\documentclass[fleqn,usenatbib]{gji}

\usepackage{newtxtext,newtxmath}
\usepackage[T1]{fontenc}
\usepackage{ae,aecompl}

\usepackage{amsmath}
\usepackage{epic}
\usepackage{graphicx}
\usepackage{hyperref}
\usepackage{multirow}
\usepackage{natbib}
\usepackage{overpic}
\usepackage{color}
\usepackage{mathtools}
\usepackage{orcidlink}
\newcommand{\norm}[1]{\lVert#1\rVert}

\AtBeginShipout{%
 \ifnum\value{page}>1 %
 \typeout{* Additional boxing of page `\thepage'}%
 \setbox\AtBeginShipoutBox=\hbox{\copy\AtBeginShipoutBox}%
 \fi
}
\title[Seismic signals generated from SPH simulations]{Seismic signals generated at a water-ice interface from smoothed particle hydrodynamic simulations}
\author[Turner et al.]{Ross J. Turner$^{\,\orcidlink{0000-0002-4376-5455}\,1\thanks{turner.rj@icloud.com}}$, Jared C. Magyar$^{\,\orcidlink{0000-0002-2655-9895}\,1}$, Sue Cook$^{\,\orcidlink{0000-0001-9878-4218}\,2}$ and Anya M. Reading$^{\,\orcidlink{0000-0002-9316-7605}\,1,3}$\\
  $^1$ School of Natural Sciences (Physics), University of Tasmania, Hobart, Private Bag 37, TAS 7001, Australia \\ $^2$ Australian Antarctic Program Partnership, Institute for Marine and Antarctic Studies, University of Tasmania, Hobart, Tasmania \\ $^3$ Australian Centre of Excellence in Antarctic Science, University of Tasmania, Hobart, Australia
  }

\let\leqslant=\leq

\begin{document}

\label{firstpage}

\maketitle
\begin{abstract} % 500 word max
% Why we are doing this research
The field of ice sheet, ice shelf and glacier-related seismology, cryoseismology, has seen rapid development in recent years. As concern grows for the implications of change in the great ice sheets of Greenland and Antarctica, so instrument advances and international field programs have expanded the availability of passive seismic datasets that contain records of seismic disturbances generated by glacier processes. Some of these processes, such as basal slip and crevasse propagation, have mechanisms with plate tectonic deformation counterparts, however, many glacier-related seismic signals are generated by melt water.  A need exists, therefore, to better understand the seismic response of moving water on glaciers which can occur as continuous processes or as transient events.    
% What we did 
We present an analytic framework to model seismic signals at the interface between moving water and ice based on either new or existing smoothed particle hydrodynamic (SPH) simulations. The framework is tailored to understanding the seismic body wave response of transient moving water events and produces a simulated dataset of water-ice collisions that can be used to predict the motion due to body waves at the location of a simulated seismometer at any location on the glacier. The synthetic acceleration time series generated are invariant to the resolution of the hydrodynamic simulation, and consider frequency-dependent weak dispersion and attenuation due to the propagating medium.
% Strengths, findings
We illustrate the capability of our framework using end-member cases of water flow: the breaking of a supraglacial melt water lake dam with the subsequent flow of water over the surface of the model glacier in channels with differing geometries. We find that the different channel geometries result in waveform envelope shapes with different first and later amplitude peaks matching initial and subsequent collisions of the melt water surge with the supraglacial channel walls. The change in waveform character with distance is also captured such that the character attributes due to the process and the those due to the propagation effects may be understood. Our focus is on the waveform attributes (features) rather than exact waveform matching, in view of the dynamic nature of the glacier environment central to ongoing applications. 
% Applications
The flexibility inherent in the computational framework will allow for the simulation of the seismic signals generated by high-energy collisions in a variety of different water flow geometries including simple 3D paths into and through a glacier. We make the code available as an open source resource for the polar geophysics community with the aim of adding to the toolbox of available approaches to inform the potential future monitoring of melt water movement and related glacier processes. 

\end{abstract}


\section{Introduction}
% Para 1, Cryoseismology in general, mostly tectonic type events.
\subsection{Cryoseismology}
Cryoseismology encompasses the study of glaciated locations, regions and glacier processes using the techniques of seismology. As with seismological studies of the solid Earth, this sub-discipline includes the use of seismic waves as a means of inferring hidden structure, and providing insights on mechanisms of deformation (in its broadest sense) that generate seismic signals \citep{podolskiy2016cryoseismology, Aster_2017}. Seismic sources arising from glaciers and their near environment are extremely diverse  \citep{Walter+2015Usingglacierseismicity}, ranging from stick-slip events \citep[e.g.][]{pratt+2014, guerin2021frictional} to crevasse generation and propagation \citep[e.g.][]{roosli2014sustained}. Many such event types have solid Earth counterparts including thrust tectonic mechanisms and fault inception and growth, hence, some understanding of seismic signal generation for these glacier-related processes is possible based on established approaches of earthquake seismology \citep[e.g.][]{stein_wysession+2003textbook}. 
Given the impact of global change on the polar regions \citep{Fox-Kemper+2021IPCC} the understanding of seismic signals without tectonic counterparts, such as those relating to water movement on the margins of great ice sheets, is also of considerable importance. 

In this contribution, we aim to add to the understanding of how seismic body wave signals are generated by moving water, and the character and variability of such signals, with a focus on transient (intermittent, short-duration) events in the glacier environment. 
%The wave pattern produced by water flow in the four channel geometries is of interest given the increasing deployment of seismic arrays on active glaciers.

% Para 2, Moving water type seismology, e.g. river literature
\subsection{Seismic signals generated by moving water}
\label{sec:Moving water}
%Para 1
Sources of seismic signals related to moving water in both glacier and non-glacier environments have many variants and may be divided into continuous (long-duration) processes and transient (intermittent, short-duration) events. Sources may be further classified into those solely due to the action of water, those due to the presence of a sediment load, and those due to the geometry of a conduit.

%Para 2
Mechanisms for the production of seismic signals caused by long-duration, effectively continuous processes, concerning solely the action of water include turbulent flow in rivers \citep[][]{gimbert2014physical} (the signal caused by turbulent flow, thus estimated, can be removed enabling the changing sediment bedload of the river to be estimated using seismic data). Studies relating to the impact of storms and flood events on the seismic signature of rivers \citep[e.g.][]{hsu2011seismic, schmandt2013multiple} have also been carried out to characterise the frequency and amplitude of the resulting seismic records.

%Para 3
Transient processes produce impulsive signals that spread into complex but relatively short duration events. The body waves produced by such events are the focus of the computational framework introduced in this work, although we envision that the scope will be expanded in future studies to include surface waves as these are expected to be of significant amplitude \citep[][]{Walter+2015Usingglacierseismicity}. Glaciers and ice shelves produce short duration events due to crevasse propagation, which may be hydraulically forced \citep{Hudson+2020}. The separation of events due to transient water motion and events due to cracking will be desirable into the future. 

%Para 4
Continuous processes produce tremor-like signals, which may include signal generation mechanisms related to sediment transport \citep[e.g.][]{tsai2012physical}. Further signal generation mechanisms relate to the way in which the geometry of the conduit produces resonance effects, as in the case of a moulin \citep[e.g.][]{roosli2014sustained, roeoesli2016seismic}. The potential interplay between multiple mechanisms is also of note, for example, stick-slip mechanisms and the presence of melt water \citep[][]{Roeoesli+2016Meltwaterinfluences}. 

%Para5
Given the dynamic environment of a moving glacier, we do not envision an exact matching of waveforms for a water movement event (as one might carry out to understand a large tectonic event), but rather to use the modelling framework in the current work to inform the diversity of characteristic waveform attributes (features). The characterisation of event types using waveform attributes facilitates data-driven analysis and semi-automated workflows \citep[e.g.][]{provost2017automatic, turner2021obspy}.

\subsection{Hydrodynamic simulations}
\label{sec:Hydrodynamic simulations}

Computational fluid dynamics is the application of the fundamental laws of fluid dynamics to an $N$-dimensional grid or a set of discrete particles. These hydrodynamic simulations are applied in a range of disciplines and for different fluid types, including aeronautical engineering (air), astrophysics (plasma), and nautical and environmental engineering (water). The modelled fluid determines the preferred numerical scheme: relativistic astrophysical plasmas are well-modelled using a numerical grid with shock-capturing methods, whilst liquids may be more accurately simulated using discrete particles \citep[e.g.][]{Monaghan+2005,Valizadeh2015}.

The foremost fluid dynamics laws, regardless of the simulation type, are those describing the conservation or continuity of mass, momentum and energy. The bulk motion of Newtonian fluids is described by the Navier-Stokes equations, or the simpler Euler equations for non-viscous fluids; these are second-order partial differential equations for the time and spatially-dependent behaviour of the fluid properties (e.g. velocity, density and pressure). The fluid equation of state must also be provided to relate the response of the fluid properties to a perturbation in any other quantity; e.g. the ideal gas law relates changes in pressure, density and temperature.
Prior to running a hydrodynamic simulation the following processing steps are, in general, required: (1) the fluid is divided into grid cells (a mesh) or represented as initially uniformly-spaced discrete particles; (2) the equations of fluid motion and state are defined; and (3) the initial and boundary conditions of the fluid are defined. 
The simulation proceeds by incrementing forward in time and solving the conservation equations at each grid cell or particle at this time step. This process is repeated iteratively, generally with an adaptive step size $dt$, to run the simulation until the desired end time. The underpinning partial differential equations are functions of the fluid properties in the neighbouring grid cells/particles, and, as a result, the simulation becomes computationally intensive with increasing spatial resolution. 

The overview presented above applies equally to grid and particle-based hydrodynamic simulations. However, a notable difference resulting from the discrete nature of particles, and their non-static locations, is the addition of a \textit{smoothing kernel} in the fluid equations of particle-based simulations. In particular, the value of some scalar attribute of the fluid, $f$, at the vector position $\boldsymbol{r}$ is given by the interpolation integral \citep[e.g.][]{Gesteira+2010}
\begin{equation}
f(\boldsymbol{r}, t) = \int_{\mathcal{R}} f(\boldsymbol{r}', t)\!\; W(|\boldsymbol{r} - \boldsymbol{r}'|, \epsilon)\!\; d^{3\!}\boldsymbol{r}' \,,
\end{equation}
where $d^{3\!}\boldsymbol{r}$ is the differential for a volume integral, and $W(..., \epsilon)$ is an appropriately normalised kernel function of smoothing length $\epsilon$. The standard choice for this function is a distribution that is approximately Gaussian in shape and truncated at the 2-3$\sigma$ level. This interpolation integral is approximated using a Riemann sum over the discrete locations of the particles, $\boldsymbol{r}'$, each with known value of the scalar attribute $f(\boldsymbol{r}', t)$. The kernel is chosen as a differentiable function permitting the gradient of $f$ and divergence of $\boldsymbol{f} \equiv (f_1, f_2, f_3)$ to be calculated analytically \citep[e.g. Equations 16 and 17 of][]{Gesteira+2010}. The conservation equations describing the fluid motion can thereby be transformed to consider particles rather than continuous grid cells. 

The modelling of fluid flows in industry applications, especially when using smoothed particle simulations, is typically performed using specialised software for the use case (e.g. wind tunnel, ocean waves). In contrast, open source codes are typically simpler but can be more readily adapted for different scenarios. In this work, we use the open source \textit{PySPH} framework \citep{Ramachandran+2020}, implemented in \textit{Python} (chosen for consistency with the \textit{obspy} project; widely used for handling observational seismology data), to manage the particles, smoothing kernel, and numerical integration for the smoothed particle hydrodynamic (SPH) simulations in this work.
The \textit{PySPH} framework applies second-order predictor-corrector integrators to advance the time step, updating the density, velocity and position based on the time-derivatives in the continuity equation (conservation of mass) and the equation of motion (conservation of momentum); the position of each particle is updated using the velocity derived from the momentum equation. Meanwhile, the pressure at each time step is calculated from the equation of state using the updated value of the density. 

%%
% The velocity dispersion and attenuation of seismic waves in glacier ice depend on the frequency- and temperature-dependent properties of the material; these are collectively referred to as ``anelasticity'' \citep[e.g.][and references therein]{McCarthy+2011}.
%Using techniques to constrain the five interatomic potentials, $c_{ij}$, \citep{Nye+1957} in the hexagonal crystal structure, known as Ice 1$h$ \citep[e.g.][]{Neumeier+2018}. 
%\citet{Jona+1952} constrained all $c_{ij}$ values for solid ice at 257\,K using diffraction pattern observations from vibrating single crystals; other authors followed using techniques including measuring the resonant frequencies of monocrystalline ice \citep{Bass+1957} and sound velocity measurements, generally in the MHz-GHz frequency range \citep[e.g.][]{Proctor+1966, Dantl+1969}. 
% The excitation of the interatomic potentials in ice crystals dominates the macroscopic properties of glacier ice (and other hexagonal crystal structures) at high-frequencies. However, for ultrasonic
% \footnote{We assume the forced harmonic oscillator is operating at a frequency significantly below the resonant frequencies associated with the interatomic potentials.}

\subsection{Properties of glacier ice}
\label{sec:Seismic waves}
As a foundation for modelling the propagation of seismic waves caused by high-energy collisions of water at a water-ice interface, we now review properties of glacier ice including elastic moduli, dispersion relations and attenuation (a more detailed account is provided in Appendix \ref{sec:Seismic waves in glacier ice}).

The physical properties of glacier ice have been investigated at the scale of the crystal lattice, and also on a macro scale. The five interatomic potentials, $c_{ij}$, in the hexagonal crystal structure have been constrained using diffraction patterns for vibrating single crystals of artificial ice \citep[e.g.][]{Jona+1952, Nye+1957,Neumeier+2018}. Others techniques have been employed that include measuring the resonant frequencies and sound velocity of monocrystalline ice, generally in the MHz-GHz frequency range \citep[e.g.][]{Bass+1957, Proctor+1966, Dantl+1969}. Meanwhile, \citet{Bogorodskiy+1964} made the first measurement of the full set of $c_{ij}$ values for \emph{natural} ice extracted from Lake Ladoga, Russia.
Brillouin spectroscopy has since been used to accurately constrain the five interatomic potentials in both artificial and natural ice across a broad range of temperatures \citep[e.g.][]{Gammon+1983}. The temperature dependence of the elastic moduli ($E$, derived directly from the $c_{ij}$ terms) is well approximated by a linear function for temperatures normally encountered in glacier environments as follows  \citep[][their Equation 9]{Gammon+1983}:
\begin{equation}
E(T) = E(T_\text{ref})\frac{1 - aT}{1 - aT_\text{ref}} \,,
\end{equation}
where the temperature, $T$, and reference temperature, $T_\text{ref}$, (with known elastic modulus) are in degrees Celsius, and $a = 1.418\times10^{-3}\rm\, /^{\circ} C$.

For waves such as seismic disturbances, with frequencies below $\sim$\,$10^5$\,Hz, the polycrystalline structure of ice permits slipping along the boundary of ice crystals \citep[e.g.][]{McCarthy+2011}; this grain boundary sliding can be modelled as a forced harmonic oscillator with amplitude $\delta l \propto F_0/\xi$, for applied force $F_0$ and wave frequency $\xi$. The deformation (strain) in response to an applied force is therefore expected to increase rapidly towards lower frequencies, significantly lowering the elastic moduli of the macroscopic crystal structure. \citet{Traetteberg+1975} measured the Young's elastic modulus in both artificial and natural glacial ice at low frequencies (0.0001 to 10\,Hz) by considering the relaxation time of bulk samples upon the application of a compressive load. \citet{Nakaya+1958} measured Young's modulus for natural polycrystalline ice samples taken from an ice tunnel in Tuto, Greenland, considering the resonant frequencies of the macroscopic crystal structure in the 100-1000\,Hz range. Grain size in polycrystalline materials is known to greatly affect their elastic behaviour \citep[e.g.][]{McCarthy+2011}; for example, \citet{Traetteberg+1975} find up to a factor of two difference between artificial and naturally formed ice. The Young's modulus, $E$, informs the frequency and temperature dependence of the longitudinal (P) and transverse (S) phase velocities as follows:
\begin{subequations}
\begin{equation}
v_{p}(\xi, T) = \bigg[\frac{1 - \sigma}{(1 + \sigma)(1 - 2\sigma)} \frac{E(\xi, T)}{\varrho(T)} \bigg]^{1/2}
\label{a1}
\end{equation}
\vspace{-8pt}
\begin{equation}
v_{s}(\xi, T) = \bigg[\frac{1}{2 (1 + \sigma)} \frac{E(\xi, T)}{\varrho(T)} \bigg]^{1/2} \,,
\label{a2}
\end{equation}
\end{subequations}
where $\sigma$ is Poisson's ratio and $\varrho$ is the density of glacier ice. The value of Poisson's ratio is informed by field measurements of the phase velocity of S and P waves \citep[e.g.][and many others]{Kohnen+1973}.
%We present empirical equations for the Young's modulus of their naturally formed ice as a function of temperature and the seismic wave frequency in Appendix \ref{sec:Seismic waves in glacial ice}.
% results differ between granular ice because grain size matters

Energy losses in the propagation of seismic waves are attributed to kinetic and static friction at the grain boundaries in the glacial ice \citep[][i.e. damped oscillation]{White+1966}. \citet{Attewell+1966} found the attenuation constant in general materials is linearly proportional to frequency over eight orders of magnitude ($0.01 < \xi < 10^6\rm\, Hz$). The attenuation of longitudinal seismic waves, more specifically, was investigated by \citet{Kohnen+1971} using seismic refraction measurements near Byrd Station in West Antarctica; they measure the phase velocity and decay in wave amplitude along three profiles angled at $60^\circ$ to one another over the 80-180\,Hz frequency band. However, their value for the attenuation constant at 100\,Hz is a factor of two lower than other authors for the Antarctic and Greenland ice sheets \citep[e.g.][]{Robin+1958, Kohnen+1969}. The attenuation constants for S and P seismic waves are known to differ significantly through experiments in various material \citep[e.g.][]{McDonal+1958}. The S wave is attenuated much more rapidly due to its lower phase velocity and thus greater interaction with ice-grain boundaries per unit length travelled. As a result, the attenuation constant, $\alpha$, is expected to increase approximately in inverse proportion to the phase velocity, $v$ (i.e. $\alpha \propto 1/v$). %; this approximation is confirmed by direct measurements of S and P wave phase velocities and attenuation constants in Pierre shale \citep{McDonal+1958}. 
We synthesise the above empirical relationships in Appendix \ref{sec:Seismic waves in glacier ice} to support the values used in this work. That is, the frequency and temperature dependence of (1) Young’s modulus in naturally formed ice, (2) the phase velocity of S and P waves in the Antarctic and Greenland ice sheets, and (3) the attenuation of S and P seismic waves.

\section{Seismic wave generation and propagation}
\label{sec:Seismic wave generation and propagation}

We present a framework to model the seismic wavefield arising from fluid particle collisions against a solid interface with known incident and reflected momentum vectors. The fluid particles exert a force on the interface over a short duration informed by hydrodynamic experiments of water droplet collisions (Section \ref{sec:Droplet collisions with the water-ice interface}). The magnitude of the applied force on the interface during the collision is modelled as a damped simple harmonic oscillator, with the interface continuing to resonate after the collision (Section \ref{sec:Collision as a damped harmonic oscillator}). The imparted force is Fourier transformed into the frequency domain to consider frequency-dependent attenuation and weak dispersion in the resultant, inverse-Fourier transformed seismic signals (Section \ref{sec:Wave generation at a water-ice interface}). Longitudinal (P) and transverse (S) seismic waveforms are synthesised by considering the integral of all fluid particle collisions on the interface (Section \ref{sec:Seismic waves from ensemble of collisions}).

%In this work, we simulated the momenta of fluid particles using a smoothed particle hydrodynamics simulation.

%Fluid particles impacting on the ice-water interface will largely transfer their momentum in a direction normal to the interface with a (typically) smaller component along the plane of the interface.

\subsection{Fluid particle collisions with water-ice interface}
\label{sec:Droplet collisions with the water-ice interface}

The collision of fluid particles with a water-ice interface is modelled considering the physics of individual water droplets impacting a solid surface. Such impacts have been studied extensively using high-speed cameras, and their physics characterised in terms of the incident velocity and droplet size \citep{Engel+1955}. The force $\boldsymbol{F}(t)$ imparted on the solid interface during a water droplet collision of duration $\tau$ is described by the water-hammer equation. Specifically, the time-averaged force is given as follows:
\begin{equation}
\begin{split}
\bar{\boldsymbol{F}} &= \frac{1}{\tau} \int_{t_0}^{t_0 + \tau} \boldsymbol{F}(t) dt \\
&= - \kappa \rho c_s (\boldsymbol{v}_{\!\;\!\tau}-\boldsymbol{v}_{0}) \delta A \,,
\end{split}
\label{hammer}
\end{equation}
where $\boldsymbol{v}_{0}$ and $\boldsymbol{v}_{\!\;\!\tau}$ are the incident and reflected velocity vectors at times $t_0$ and $t_0 + \tau$ respectively, $\rho$ is the local fluid density, $c_s$ is the local sound speed, and ${\delta\!\;\!A}$ is the cross-sectional area of the collision. 
The standard water-hammer equation is modified here based on the results of hydrodynamic experiments by a constant of proportionality $\kappa$; for an isolated spherical droplet $\kappa \approx 0.2$ increasing to $\kappa = 1$ for a droplet incased in a fluid column \citep[][]{Engel+1955}. We model the time-varying behaviour of the applied force, $\boldsymbol{F}(t)$, as a damped harmonic oscillator in the following section.

The collision duration is derived considering the impulse-momentum theorem (i.e. $\bar{\boldsymbol{F}} \tau = \Delta \boldsymbol{p}$ for momentum vector $\boldsymbol{p}$), yielding the following expression for the time-averaged applied force on the interface:
\begin{equation}
\bar{\boldsymbol{F}} = -\frac{\rho \delta V (\boldsymbol{v}_{\!\;\!\tau}-\boldsymbol{v}_{0})}{\tau} \,.
\label{impulse}
\end{equation}
where $\delta V$ is the fluid particle volume. Comparing these two expressions for the applied force on the interface (Equations \ref{hammer} and \ref{impulse}) gives the collision duration as,
\begin{equation}
\tau = \frac{\delta V}{\kappa c_s \delta A} \,,
\label{tau}
\end{equation}
where for a sound speed of $c_s = 1400\rm\, m\,s^{-1}$ (for water at $0^\circ$C and standard atmospheric pressure), the duration of the collision is between $\tau \approx 0.0007\!\;\!\, \delta l$ and $0.0036\!\;\!\, \delta l$\,s for fluid particle diameter $\delta l$ (considering the plausible range in $\kappa$).

The collision duration is therefore expected to be shorter than the sampling interval of a typical cryoseismic survey (i.e. 0.001\,s for 1000\,Hz; 0.005\,s for 200\,Hz) for fluid particles with diameter $\delta l < 0.28$\,m; this can be achieved for even the coarsest resolution hydrodynamic simulations. As a result, the applied force resulting from simulated fluid particle collisions will be averaged within a single temporal sample (or split over two neighbouring samples) of duration $\delta t \approx 0.001$-0.005\,s. Seismic signals generated based on the impulse of fluid particle collisions are therefore invariant to the simulation resolution, at least of the order considered in this work (i.e. $\delta l = 0.01$\,m; see Section \ref{sec:Smoothed particle hydrodynamic simulations}).
%The stability of our modelled seismic signals to small changes in the assumed collision duration is tested in Section \ref{sec:discussion} as the value of this parameter is expected to vary somewhat with the simulation resolution, and to a lesser extent depending on the geometry of the colliding droplet.

\subsection{Particle collisions as damped harmonic oscillators}
\label{sec:Collision as a damped harmonic oscillator}

We model the time-varying force, $\boldsymbol{F}(t)$, applied on a water-ice interface during the $n$th fluid particle collision as a damped simple harmonic oscillator.
The force applied by the incident fluid particle will accelerate the ice at the impact site leading to oscillations and the generation of a seismic wave. The location of the $n$th fluid particle collision is defined by the position vector $\boldsymbol{r}_{\!\;\!n} = (x_{n},y_{n},z_{n})$, where we assume Cartesian coordinates in this work.
The force exerted on the ice by the $n$th fluid particle collision (for $t_n \leqslant t \leqslant t_n + \tau_n$) is related to its Cartesian acceleration as follows:
\begin{equation}
\boldsymbol{F}_{\!\;\! n}(\boldsymbol{r}_{\!\;\!n}, t) = -\rho_{n} \delta V_n \boldsymbol{a}_n(\boldsymbol{r}_{\!\;\!n}, t) \,,
\label{force}
\end{equation}
where $\rho_{n}$ and $\delta V_n$ are the density and volume of the $n$th fluid particle respectively (the product $\rho_{n} \delta V_n$ remains constant for each particle in SPH simulations), and $\boldsymbol{a}_n(\boldsymbol{r}_{\!\;\!n}, t)$ is the acceleration of the fluid particle during the $n$th collision. 
The form of this acceleration vector during the collision is constrained by considering: (1) the kinetic energy, then (2) the displacement of the water-ice interface before and after the collision. The amplitude, direction and damping of the acceleration vector for each fluid particle collision is constrained based on incident and reflected velocity vectors output from hydrodynamic simulations (Section \ref{sec:Smoothed particle hydrodynamic simulations}).

\subsubsection{Kinetic energy during collision}

The time-dependence of the acceleration of a fluid particle during its interaction with the water-ice interface is derived assuming an inelastic collision, and as such, we assume some fraction of the incident kinetic energy is lost during the collision. 
The ratio of the kinetic energy after and before the collision of the $n$th fluid particle is given as follows:
\begin{equation}
\begin{split}
\frac{K_{\tau,n}}{K_{0,n}} = \frac{\norm{\boldsymbol{v}_{\!\;\!\tau,n}}}{\norm{\boldsymbol{v}_{0,n}}} = \exp\!\;\!\left[- 2 \omega_{n} \zeta_n (1 - \zeta_n^2)^{-1/2} \tau_n \right] \,,
\end{split}
\label{energy}
\end{equation}
where $\boldsymbol{v}_{0,n}$ and $\boldsymbol{v}_{\!\;\!\tau,n}$ are the incident and reflected velocity vectors of the $n$th fluid particle. Meanwhile, in the second equality, the kinetic energy ratio is expressed in terms of damping coefficients; i.e. $\omega_{n}$ and $\zeta_n$ are the (damped) angular frequency and damping ratio of the $n$th fluid particle collision respectively. 

Most fluid particles will, in practice, have some (potentially sizeable) component of their initial kinetic energy that is not involved in the collision; e.g. the velocity component parallel to the water-ice interface is not converted into elastic potential energy. We therefore remove this residual velocity component, $\boldsymbol{v}_{\!\;\!\star,n}$ (i.e. the vector at the point of minimum kinetic energy), from the incident and reflected velocity vectors (see Section \ref{sec:Smoothed particle hydrodynamic simulations}). The velocity vectors in Equation \ref{energy} are therefore rewritten as $\boldsymbol{v}_{0,n} - \boldsymbol{v}_{\!\;\!\star,n}$ and $\boldsymbol{v}_{\!\;\!\tau,n} - \boldsymbol{v}_{\!\;\!\star,n}$ respectively.

The damping ratio of the $n$th fluid particle collision is derived by comparing the above two expressions for the kinetic energy ratio after and before its interaction with the water-ice interface. That is,
\begin{equation}
\begin{split}
\zeta_{n} = \left(\frac{(\omega_n \tau_n)^2}{\log^2 \!\;\! \big(\norm{\boldsymbol{v}_{0,n} - \boldsymbol{v}_{\!\;\!\star,n}}/\norm{\boldsymbol{v}_{\!\;\!\tau,n} - \boldsymbol{v}_{\!\;\!\star,n}}\big)} + 1 \right)^{-1/2} \,,
\end{split}
\label{damping}
\end{equation}
where we assume that the kinetic energy of the reflected fluid particle is less than that of the incident particle. 
We have now constrained the decay envelope of the oscillation, but must consider the displacement at the water-ice interface to fully define the fluid particle behaviour during the collision.

%The fluid particle will be reflected upon interacting with the boundary, and consequently, the component of the velocity vector normal to the interface must change sign during the collision; i.e. $(\boldsymbol{v}_{0} \cdot \hat{\boldsymbol{n}})(\boldsymbol{v}_{\!\;\!\tau} \cdot \hat{\boldsymbol{n}}) < 0$ for unit vector $\hat{\boldsymbol{n}}$ normal to the water-ice interface. By contrast, a large fraction of the kinetic energy associated with velocity components parallel to the interface may not involved in the collision; e.g. the particle velocity is only reduced in magnitude rather than changing sign. As such, we define any non-zero component of the fluid particle velocity at its maximum displacement from equilibrium as the bulk flow velocity, $\boldsymbol{v}_{\!\;\! b, n} = (v_{b x,n}, v_{b y,n}, v_{b z,n})$. The incident and reflected velocity vectors obtained from our smoothed particle hydrodynamic simulation outputs have this bulk flow velocity component removed (see Section \ref{sec:Smoothed particle hydrodynamic simulations}); this yields a sign change in the velocity components parallel to the interface.

\subsubsection{Displacement during collision}
\label{sec:Displacement during collision}

The motion of the fluid particle during the $n$th collision is represented by a damped harmonic oscillation with angular frequency $\omega_n$ and damping ratio $\zeta_n$, as discussed above. The displacement of the fluid particle from the equilibrium position of the collision site, along the some arbitrary axis $q$, is defined as $\eta_{q,n} \equiv q - q_n$. The time-varying displacement from equilibrium is thus given by,
\begin{equation}
\begin{split}
\eta_{q,n}(t) &= \eta_{0q,n} \exp\!\;\!\left[{-\omega_{n} \zeta_{n} (1 - \zeta_n^2)^{-1/2} (t - t_n)}\right]\\  
&\!\!\!\!\!\!\!\!\times \sin\!\;\!\left[\omega_{n} (t - t_n) + \phi \right] + v_{\star q,n} (t - t_n),
\end{split}
\label{eta_x}
\end{equation}
where $\eta_{0q,n}$ is the amplitude of the $q$-component of the oscillation in the fluid particle during the collision, $t_n$ is the time at the start of the collision, and $\phi$ is an arbitrary phase constant. The $q$-component of the residual velocity vector, $v_{\star q,n}$ (i.e. the velocity component not energetically involved in the collision), shifts the displacement linearly with time from the equilibrium position.
The velocity of this $n$th fluid particle during its collision with the water-ice interface is similarly given by,
\begin{equation}
\begin{split}
v_{q,n}(t) &= \eta_{0q,n} \omega_n \exp\!\;\!\left[{-\omega_{n} \zeta_{n} (1 - \zeta_n^2)^{-1/2} (t - t_n)}\right]\\ 
&\!\!\!\!\!\!\!\!\times \Big[ - \zeta_{n} (1 - \zeta_n^2)^{-1/2}\sin\!\;\!\big(\omega_{n} (t - t_n) + \phi\big)\\
&\!\!\!\!\!\!\!\!\quad\quad\quad\quad\quad\quad + \cos\!\;\!\big(\omega_{n} (t - t_n) + \phi \big)\Big] + v_{\star q,n} \,.
\end{split}
\label{v_x}
\end{equation}

The fluid particle will be reflected upon interacting with the ice boundary, and consequently, the component of the velocity vector normal to the interface is expected to change sign during the collision; i.e. $(\boldsymbol{v}_{0,n} \cdot \hat{\boldsymbol{n}})(\boldsymbol{v}_{\!\;\!\tau,n} \cdot \hat{\boldsymbol{n}}) < 0$ for unit vector $\hat{\boldsymbol{n}}$ normal to the interface. The initial condition, at time $t = t_n$, for this normal-component of the displacement of the incident fluid particle is $\eta_{\perp, n}(t_n) = 0$; this sets the arbitrary phase constant as $\phi = 0$. After the collision, at $t = t_n + \tau_n$, the displacement of the fluid particle is again $\eta_{\perp,n}(t_n + \tau_n) = 0$ assuming that $v_{\star \perp,n} = 0$ (though see below). The angular frequency of the oscillation is thus $\omega_{n} = m \pi/\tau_n$ for integer $m$. The displacement of the fluid particle normal to the interface is logically modelled as a half-period oscillation to yield the expected sign change in the velocity vector; i.e. $m=1$ and $\omega_{n} = \pi/\tau_n$. 

The motion parallel to the interface must similarly be modelled as a half-period damped harmonic oscillation, noting a single angular frequency and damping ratio were assumed when considering the kinetic energy. The displacement before and after the collision is $\eta_{\parallel, n}(t_n) = 0$ and $\eta_{\parallel,n}(t_n + \tau_n) = v_{\star\parallel,n} \tau_n$ respectively, where $v_{\star\parallel,n}$ is the potentially sizeable component of the velocity vector parallel to the boundary that is not energetically involved in the collision. Regardless, these conditions provide the same constraints on the parameters in our damped harmonic oscillation model.

The initial condition on the kinetic energy of the fluid particle can therefore be used to set the displacement amplitude of the oscillation. That is, comparing the measured velocity vector in Equation \ref{energy} and the expression for the oscillator velocity in Equation \ref{v_x}, we find that $\norm{\boldsymbol{\eta}_{0,n}} = \norm{\boldsymbol{v}_{0,n} - \boldsymbol{v}_{\!\;\!\star,n}}/\omega_n$. Modelling the motion of the fluid particle would additionally require the displacement vector to be separated into Cartesian components, however, we are only concerned with the direction of the applied force, as we now describe.

%We assume the collision can be modelled as a half-period harmonic oscillation (i.e. $m = 1$) as this will yield the expected sign change in the component of the velocity vector normal to the interface; i.e. $(\boldsymbol{v}_{0} \cdot \hat{\boldsymbol{n}})(\boldsymbol{v}_{\!\;\!\tau} \cdot \hat{\boldsymbol{n}}) < 0$ for unit vector $\hat{\boldsymbol{n}}$ normal to the water-ice interface. However, a large fraction of the kinetic energy associated with velocity components parallel to the interface may not involved in the collision; at first pass, our initial and final conditions do not see consistent with this scenario.

\subsubsection{Time-varying force during collision}

The direction of the time-averaged applied force (Section \ref{sec:Droplet collisions with the water-ice interface}) was derived considering the water-hammer equation and impulse-momentum theorem. Here, we will assume the time-varying force is oriented in this same direction but varies in magnitude throughout the duration of the collision; i.e. the $q$-axis is in the $\boldsymbol{v}_{\!\;\!\tau,n} - \boldsymbol{v}_{0,n}$ direction.
The magnitude of the corresponding acceleration on the $n$th fluid particle during the collision is found from the derivations in Section \ref{sec:Displacement during collision}. That is,
\begin{equation}
\begin{split}
{a}_{q,n}(t) &= -\omega_n \frac{\norm{\boldsymbol{v}_{0,n} - \boldsymbol{v}_{\!\;\!\star,n}}}{(1 - \zeta_{n}^2)} \left( 2\zeta_{n}(1 - \zeta_{n}^2)^{1/2} + i(2\zeta_{n}^2 - 1) \right) \\
&\!\!\!\!\!\!\!\! \times \exp\!\;\!\left[{-\omega_n \zeta_{n}(1 - \zeta_{n}^2)^{-1/2} (t - t_n)}\right] \exp\!\;\!\big[i\omega_n (t - t_n)\big] \,,
\end{split}
\label{a_x1}
\end{equation}
where the complex-valued phase is included for the derivations presented in the following section. 

The time-varying force exerted on the ice during the collision is proportional to this acceleration for $t_n \leqslant t \leqslant t_n + \tau_n$ (see Equation \ref{force}). The water-ice interface will continue to oscillate after the collision due to restoring forces within the deformed ice. However, before the collision, there can clearly be no force applied on the ice by the fluid particle. 
The time-varying force experienced by the water-ice interface due to the $n$th fluid particle collision, and subsequent internal restoring forces, is given at all times (before, during and after the collision) as follows
\begin{equation}
\boldsymbol{F}_{\!\;\!n}(\boldsymbol{r}_{\!\;\!n}, t) = -\rho_n \delta V_n a_{q,n}(t) u(t - t_n) \frac{(\boldsymbol{v}_{\!\;\!\tau,n} - \boldsymbol{v}_{0,n}) }{\norm{\boldsymbol{v}_{\!\;\!\tau,n} - \boldsymbol{v}_{0,n}}} \,,
\label{a_x2}
\end{equation}
where $u(t - t_n)$ is the Heaviside (unit step) function, required as the force is zero at times before the impact, and the unit vector for the $q$-axis is now included.

%The impacted volume of ice is accelerated at the same rate as the fluid particle in response to the collision; this is a statement of continuity of number density across the water-ice interface during the collision. As a result, from Equation \ref{force}, the volume of ice directly impacted by the fluid particle is $\delta V = \rho_n \delta V_n/\rho$; as the cross-sectional area of the collision in both the water and ice is $(\delta V_n)^{2/3}$, the depth of the forced volume of ice is $\rho_n (\delta V_n)^{1/3}/\rho$ implying that $\delta V$ is not $x$-, $y$-, $z$-axis symmetric. 
%The $y$- and $z$-components of the displacement, velocity, acceleration and damping ratio have the same form as the $x$-component.

\subsection{Generation of synthetic seismic waves}
\label{sec:Wave generation at a water-ice interface}

We now present a technique to produce synthetic seismic waveforms from the time-varying force applied on the water-ice interface by each fluid particle collision. This formalism considers frequency-dependent attenuation and weak dispersion in the seismic waves emanating from each collision site; this is vital due to the potentially broad frequency range of signals produced by water-ice collisions. The incised supraglacial flows simulated in this work are expected to primarily produce body waves, and thus we focus on this mechanism; however, the framework presented here can be readily modified to additionally consider surface waves when modelling (e.g.) supraglacial flows.

%considered in this work are assumed to be surrounded by constant density ice, and thus the seismic waves travel in straight lines, though may be reflected off the ice-bedrock interface, and to a lesser extent, the ice-surface interface.

%\subsubsection{Fourier spectrum of applied force}
\subsubsection{Weak dispersion}
\label{sec:Fourier spectrum and dispersion of the collision}

The oscillation generated by the time-varying applied force is considered after it has propagated some distance through the ice to a seismometer location (noting that, should it be required, the near-field motion would be equally-well captured as the far-field by the modelling framework).
The applied force at the water-ice interface is represented as a superposition of forces arising from monochromatic wave packets.
The applied force amplitude associated with each monochromatic wave packet is found by taking the Fourier transform of the time-varying applied force derived in Equations \ref{force} and \ref{a_x1}. That is,
%
%\begin{equation}
%\begin{split}
%A_{n}(\xi) &= \frac{\pi \norm{\boldsymbol{v}_{0,n} - \boldsymbol{v}_{\!\;\!\star,n}}}{(1 - \zeta_n^2)} \exp\!\;\! \left[-i2\pi \xi t_n \right] \\
%&\!\!\!\!\!\!\!\! \times \Bigg[ \frac{\pi(1 - 2\zeta_n^2) + 2\tau\zeta_{x,n}(1 - \zeta_n^2)^{1/2}\,(\zeta_n \omega_n + i2\pi \xi)}{\pi^2 + \tau^2(\zeta_n \omega_n + i2\pi \xi)^2}  \Bigg] \,.
%\end{split}
%\label{transform}
%\end{equation}
%
\begin{equation}
\begin{split}
\mathcal{F}_{n}(\boldsymbol{r}_{\!\;\!n}, \xi) &= \rho_n \delta V_n \frac{\norm{\boldsymbol{v}_{0,n} - \boldsymbol{v}_{\!\;\!\star,n}}}{(1 - \zeta_{n}^2)} \\
&\!\!\!\!\!\!\!\! \times \left( \frac{2\zeta_{n}(1 - \zeta_{n}^2)^{1/2} + i(2\zeta_{n}^2 - 1)}{2\pi \xi/\omega_n +  \zeta_{n}(1 - \zeta_{n}^2)^{-1/2} - i} \right) \,.
 \end{split}
\label{transform}
\end{equation}
where $\xi \geqslant 0$ is the frequency of a given monochromatic wave packet.
The dispersion of each such wave packet is modelled by considering its spatial evolution in Fourier space. The applied force amplitude of a wave packet at frequency $\xi$ after propagating a distance $\norm{\boldsymbol{r} - \boldsymbol{r}_{\!\;\!n}}$ is given by,
\begin{equation}
\begin{split}
\mathcal{F}_{n}(\boldsymbol{r}, \xi) = \mathcal{F}_{n}(\boldsymbol{r}_{\!\;\!n}, \xi) e^{- i k(\xi) \norm{\!\;\!\,\boldsymbol{r} - \boldsymbol{r}_{\!\;\!n}\!\;\!} } \,,
\end{split}
\end{equation}
where we assume the seismic wave is propagating towards the seismometer; i.e. the seismic wavevector $\boldsymbol{k}_{\!\;\!n}$ is parallel to the position vector $\boldsymbol{r} - \boldsymbol{r}_{\!\;\!n}$.
The amplitude of this frequency-dependent seismic wavevector in ice, $k(\xi)$ (i.e. the wavenumber), is defined in Equations \ref{wavenumber1} and \ref{wavenumber2} in Appendix \ref{sec:Frequency dependent wave propagation in ice}.

The applied force at the seismometer location, so far only considering weak dispersion, is then found by taking the inverse-Fourier transform of this function. That is,
\begin{equation}
\begin{split}
F_{n}(\boldsymbol{r},t) &= \frac{1}{2\pi} \int_{0}^\infty \mathcal{F}_{n}(\boldsymbol{r}, \xi) e^{i2\pi \xi (t - t_n)} d\xi \\
&= \frac{1}{2\pi} \int_{0}^\infty \mathcal{F}_{n}(\boldsymbol{r}_{\!\;\!n}, \xi) e^{i(2\pi \xi (t - t_n) - k(\xi) \norm{\!\;\!\,\boldsymbol{r} - \boldsymbol{r}_{\!\;\!n}\!\;\!})} d\xi \,,
\end{split}
\end{equation}
The inverse-Fourier transform cannot be solved analytically for a general dispersion relation, but can be evaluated for a linear function as we now describe. We perform a Taylor series expansion on the frequency-dependent wavenumber at some expansion point $\xi = \xi_c$ to obtain an analytically tractable solution. That is,
\begin{equation}
\begin{split}
k(\xi) \approx k(\xi_c) + (\xi - \xi_c)\frac{\partial k(\xi_c)}{\partial \xi} \,,
\end{split}
\label{k_x}
\end{equation}
where the partial derivative of the wavevector is defined in Equations \ref{wavenumberd1} and \ref{wavenumberd2} of Appendix \ref{sec:Frequency dependent wave propagation in ice}. However, the chosen expansion point will  greatly affect the accuracy of the inverse-Fourier transform. We therefore examine the magnitude of the applied force amplitude in Equation \ref{transform} as a function of frequency to determine an appropriate expansion point. That is,
\begin{equation}
\begin{split}
\norm{\mathcal{F}_{n}(\boldsymbol{r}_{\!\;\!n}, \xi)} &= \rho_n \delta V_n \frac{\norm{\boldsymbol{v}_{0,n} - \boldsymbol{v}_{\!\;\!\star,n}}}{(1 - \zeta_{n}^2)} \\
&\!\!\!\!\!\!\!\! \times \left( {1 + \big[2\pi \xi/\omega_n + \zeta_{n}(1 - \zeta_{n}^2)^{-1/2}\big]^2} \right)^{-1/2} \,.
 \end{split}
\end{equation}
This function has a global maximum (on the restricted domain $\xi \geqslant 0$) at $\xi = 0$, with a value of $\rho_n \delta V_n (1 - \zeta_{n}^2)^{-1/2} \norm{\boldsymbol{v}_{0,n} - \boldsymbol{v}_{\!\;\!\star,n}}$, decreasing as approximately $1/\xi$ with increasing frequency. The expansion point $\xi_c$ is therefore taken as a lower bound for the seismometer frequency-response function to ensure greatest accuracy for the highest amplitude monochromatic wave packets; we assume $\xi_c = 0.01\rm\, Hz$ in this work. The approximated wavenumber is accurate to within 6\%, 15\% and 24\% for frequencies up to 0.1, 1 and 10\,Hz respectively, with still higher frequencies having comparably insignificant applied force amplitudes.

The force induced by the seismic wave packet from the $n$th fluid particle can therefore be expressed as a function of time and position as follows: 
\begin{equation}
\begin{split}
\boldsymbol{F}_{n}(\boldsymbol{r},t) &= \boldsymbol{F}_{n}\!\;\! \left(\boldsymbol{r}_{\!\;\! n}, t - \tfrac{\partial k(\xi_c)}{\partial \xi}  \tfrac{\norm{\boldsymbol{r} - \boldsymbol{r}_{\!\;\!n}}}{2\pi}\right) \\
&\!\!\!\!\!\!\!\!\times \exp\!\;\! \left[-i \left(k(\xi_c) - \tfrac{\partial k(\xi_c)}{\partial \xi} \xi_c \right) \norm{\boldsymbol{r} - \boldsymbol{r}_{\!\;\!n}} \right] \,,
\end{split}
\end{equation}
where the first term on the right-hand side of the equality is $\boldsymbol{F}_{n}(...)$, as given in Equation \ref{a_x2}, evaluated at the relevant (earlier) time considering the travel time from the location of the collision. 

\subsubsection{Attenuation}
\label{sec:Attenuation and inverse-square law expansion}

The synthetic seismic waves lose energy as they propagate towards a seismometer location through two mechanisms: (1) attenuation due to static and kinetic friction between oscillating granular ice crystals (see Appendix \ref{sec:attenuation}); and (2) geometric spreading into the solid angle on the ice side of the water-ice interface. 

Firstly, the frequency-dependent attenuation vector, $\boldsymbol{\alpha}_{n}(\xi)$, is directed parallel to the position vector and, as for the wavenumber, must be linearised to obtain an analytically tractable solution to the inverse-Fourier transform. That is, the Taylor series expansion for the amplitude of the attenuation vector is given by,
\begin{equation}
\begin{split}
\alpha(\xi) \approx \alpha(\xi_c) + (\xi - \xi_c)\frac{\partial \alpha(\xi_c)}{\partial \xi} \,,
\end{split}
\label{beta_x}
\end{equation}
where the amplitude of the attenuation vector at the expansion point, $\alpha(\xi_c)$, is defined in Equations \ref{alpha1} and \ref{alpha2} of Appendix \ref{sec:attenuation}, and the partial derivative is defined in Equations \ref{alphad1} and \ref{alphad2}.

Secondly, the energy loss due to the inverse-square law divergence of the wave simply introduces a factor of $\delta A_n/(2\pi \norm{\boldsymbol{r} - \boldsymbol{r}_n}^2)$; here, $\delta A_n$ is the cross-sectional area of the $n$th fluid particle collision at the water-ice interface, and $2\pi \norm{\boldsymbol{r} - \boldsymbol{r}_n}^2$ is the surface area the seismic wave energy passes through an arbitrary distance from the collision site.

Considering these two forms of wave attenuation, in addition to the previously described weak dispersion, we again derive the applied force amplitude of the series of monochromatic wave packets at the seismometer. That is, the applied force amplitude for a wave packet at frequency $\xi$ is given by,
\begin{equation}
\begin{split}
\mathcal{F}_{n}(\boldsymbol{r}, \xi) = \mathcal{F}_{n}(\boldsymbol{r}_{\!\;\!n}, \xi) e^{-\{\alpha(\xi) + i k(\xi)\} \norm{\!\;\!\,\boldsymbol{r} - \boldsymbol{r}_{\!\;\!n}\!\;\!} } \frac{\delta A_n}{2\pi \norm{\boldsymbol{r} - \boldsymbol{r}_{\!\;\!n}}^2 } \,,
\end{split}
\label{force_transform}
\end{equation}
where $\mathcal{F}_{n}(\boldsymbol{r}_{\!\;\!n}, \xi)$ is again defined in Equation \ref{transform}.

The force induced by the seismic wave packet from the $n$th fluid particle collision, considering both weak dispersion and attenuation, can therefore be expressed as a function of time and position as follows: 
\begin{equation}
\begin{split}
\boldsymbol{F}_{\!\;\!n}(\boldsymbol{r},t) &= \boldsymbol{F}_{\!\;\!n} \!\;\! \left(\boldsymbol{r}_{\!\;\! n}, t - \tfrac{\partial k(\xi_c)}{\partial \xi}  \tfrac{\norm{\!\;\!\,\boldsymbol{r} - \boldsymbol{r}_{\!\;\!n}\!\;\!}}{2\pi} + i\tfrac{\partial \alpha(\xi_c)}{\partial \xi} \tfrac{\norm{\!\;\!\,\boldsymbol{r} - \boldsymbol{r}_{\!\;\!n}\!\;\!}}{2\pi}\right) \\
&\!\!\!\!\!\!\!\! \times 
\exp \!\;\! \left[-\left(\alpha(\xi_c) - \tfrac{\partial \alpha(\xi_c)}{\partial \xi} \xi_c \right) \norm{\boldsymbol{r} - \boldsymbol{r}_{\!\;\!n}} \right] \\
&\!\!\!\!\!\!\!\!\times \exp\!\;\! \left[-i \left(k(\xi_c) - \tfrac{\partial k(\xi_c)}{\partial \xi} \xi_c \right) \norm{\boldsymbol{r} - \boldsymbol{r}_{\!\;\!n}} \right] \frac{\delta A_n}{2\pi \norm{\boldsymbol{r} - \boldsymbol{r}_n}^2} \,.
\end{split}
\label{a_x3}
\end{equation}
We only consider the real-valued component of the resulting force throughout the remainder of this work. 

\subsubsection{Acceleration at the seismometer}
\label{sec:Displacement at the seismometer due to the collision}

Seismometers generally record a voltage proportional to either the velocity or acceleration (in the case of MEMS-type instruments) of ground motion as a function of time. We convert the force into an acceleration based on the mass of ice being displaced; subsequently, the velocity is calculated from the acceleration time series in this work.  As discussed previously (Section \ref{sec:Droplet collisions with the water-ice interface}), the applied force (and thus acceleration) is invariant to the assumed collision duration and spatial resolution of the simulation.

The acceleration of ice at the seismometer location from the $n$th fluid particle collision is given by,
\begin{equation}
\boldsymbol{a}_{n}(\boldsymbol{r}, t) = -\frac{1}{\varrho \bar{\delta V}} \boldsymbol{F}_{\!\;\!n}(\boldsymbol{r},t)\,,
\label{eta_vect}
\end{equation}
where $\varrho = 916\rm\,kg\,m^{-3}$ is the density of ice (at $T=0^\circ$C and atmospheric pressure), and $\bar{\delta V}$ is the mean (or initial) fluid particle volume.

\subsection{Ensemble of synthetic seismic waves}
\label{sec:Seismic waves from ensemble of collisions}

The collision of a fluid particle on the water-ice interface creates an oscillation at the collision site, as described in the previous section. The energy of the impact primarily radiates outwards from the disturbance through the ice as longitudinal seismic waves, but a fraction of the energy generates transverse waves due to the shear stresses in the ice; the transverse wave energy is concentrated at angles normal to the angle of incidence. As a result, the energy imparted on the water-ice interface is not expected to radiate through the ice equally in all directions if the water has a particular angle of incidence. Moreover, longitudinal and transverse seismic waves propagate through ice at different speeds and with differing rates of attenuation.

The acceleration of each small volume of ice, due to the collision of the $n$th fluid particle, must therefore be broken down into components parallel and transverse to the direction of wave propagation to correctly consider seismic wave propagation through the material.
The longitudinal (P) wave results from the component of oscillations parallel to the direction of propagation, $\boldsymbol{r} - \boldsymbol{r}_n$, defined from the impact site $\boldsymbol{r}_n$ to the seismometer $\boldsymbol{r}$. The resulting acceleration at the seismometer is given by,
\begin{equation}
\boldsymbol{a}_{\!\;\!p,n}(\boldsymbol{r},t) = \frac{\boldsymbol{a}_{n}(\boldsymbol{r},t, \alpha_p, k_p) \cdot (\boldsymbol{r} - \boldsymbol{r}_{\!\;\!n})}{\norm{\boldsymbol{r} - \boldsymbol{r}_{\!\;\!n}}^2} (\boldsymbol{r} - \boldsymbol{r}_{\!\;\!n}) \,,
\end{equation}
where the expression for the acceleration derived in Equation \ref{eta_vect} is evaluated for the magnitude of the frequency-dependent wavenumber, $k_{\!\;\!p}(\xi)$, and attenuation constant, $\alpha_{\!\;\!p}(\xi)$, for the longitudinal component of the wave as given in Appendices \ref{sec:Frequency dependent wave propagation in ice} and \ref{sec:attenuation} respectively.

Meanwhile, the transverse (S) wave is due to oscillations normal to the direction of propagation; i.e. components of the oscillation not parallel to the direction of wave propagation. The resulting acceleration induced at $\boldsymbol{r}$ is given by,
\begin{equation}
\begin{split}
\boldsymbol{a}_{s,n}(\boldsymbol{r},t) = \boldsymbol{a}_{n}(\boldsymbol{r},t, \alpha_s, k_s) - 
\frac{\boldsymbol{a}_{n}(\boldsymbol{r},t, \alpha_s, k_s) \cdot (\boldsymbol{r} - \boldsymbol{r}_{\!\;\!n})}{\norm{\boldsymbol{r} - \boldsymbol{r}_{\!\;\!n}}^2} (\boldsymbol{r} - \boldsymbol{r}_{\!\;\!n}) \,,
\end{split}
\end{equation}
where, here, we use the magnitude of the frequency-dependent wavenumber, $k_s(\xi)$, and attenuation constant, $\alpha_s(\xi)$, for the transverse component of the wave.

The total acceleration induced at the seismometer location by the collision from the $n$th fluid particle is therefore given by,
\begin{equation}
\boldsymbol{a}_{\!\;\!ps,n}(\boldsymbol{r},t) = \frac{\boldsymbol{a}_{\!\;\!p,n}(\boldsymbol{r},t) + \Lambda \!\;\!\,\boldsymbol{a}_{s,n}(\boldsymbol{r},t)}{1 + \Lambda} \,,
\end{equation}
where $\Lambda^{\!\;\!2}$ is the ratio of energy imparted into the transverse and longitudinal seismic waves; we assume a value of $\Lambda^{\!\;\!2} = 0.25$ (i.e. $\Lambda = +0.5$) throughout this work.
The seismic signal resulting from numerous fluid particle collisions at various locations, $\boldsymbol{r}_{\!\;\!n}$, along the water-ice interface is found by summing the contributions from each collision. That is,
\begin{equation}
\begin{split}
\boldsymbol{a}(\boldsymbol{r}, t) = \sum_n^{} \boldsymbol{a}_{\!\;\!ps,n}(\boldsymbol{r}, t) \,.
\end{split}
\label{ensemble}
\end{equation}
This function is sampled at a frequency of $200\rm\,Hz$ (typical for glacier seismology deployments) to produce a three-component seismic record of the ensemble of water particle collisions. The synthetic displacement and velocity time series measurements are obtained following the same methodology to facilitate comparison with seismic waveforms as they might be typically observed (see Section \ref{sec:Displacement and velocity time series}).

\section{Smoothed particle hydrodynamic simulations}
\label{sec:Smoothed particle hydrodynamic simulations}

\subsection{Fluid dynamics of water and ice}

The hydrodynamic simulations in this work involve two types of particles: (1) the fluid particles of the water; and (2) the solid ice bounding the fluid. We do not explicitly model air particles but instead set pressure of the fluid surface to a reference atmospheric pressure, $p_0$. The fluid dynamics assumed for the ice and water, including their interaction, are detailed in the following sections.

\subsubsection{Water particle conservation equations}

The bulk motion of water is modelled in this work using SPH simulations, as introduced in Section \ref{sec:Hydrodynamic simulations}. We use the open source \textit{PySPH} framework \citep{Ramachandran+2020}, as noted previously, which enables efficient management of the particles, smoothing kernel and numerical integration.
In this code, the fluid dynamics of water is captured by three conservation equations for its continuity, momentum and equation of state. 

The continuity equation relates the rate of change in density to the divergence of the fluid particle velocity as \citep[e.g.][]{Monaghan+1992, Monaghan+2005}:
\begin{equation}
\frac{d \rho(\boldsymbol{r})}{d t} = - \rho(\boldsymbol{r}) \left[\nabla \cdot \boldsymbol{v}(\boldsymbol{r}) \right] \,,
\end{equation}
where the discrete divergence operator for the particle at location $\boldsymbol{r}$ is defined in (e.g.) Equation 17 of \citet{Gesteira+2010}. 

Meanwhile, the momentum equation describes the particle acceleration in terms of the pressure gradient, fluid viscosity and external forces as:
\begin{equation}
\frac{d \boldsymbol{v}(\boldsymbol{r})}{d t} = \frac{1}{\rho(\boldsymbol{r})}  \left[ - \nabla p(\boldsymbol{r}) + \nabla \cdot \boldsymbol{\pi}(\boldsymbol{r}) \right] + \boldsymbol{g}(\boldsymbol{r}) \,,
\end{equation}
where the discrete gradient operator for the particle at location $\boldsymbol{r}$ is defined in (e.g.) Equation 16 of \citet{Gesteira+2010}, $\boldsymbol{\pi}(\boldsymbol{r})$ is the deviatoric stress tensor (viscosity terms in the Cauchy stress tensor), and $\boldsymbol{g}(\boldsymbol{r}) = -9.81 \boldsymbol{\hat{z}}\rm\,m\,s^{-2}$ is the acceleration from external forces for unit vector $\boldsymbol{\hat{z}}$ in the vertical direction. The divergence of the deviatoric stress tensor is approximated with an artificial viscosity that includes linear velocity terms that simulate shear and bulk viscosity, and a quadratic term to consider high Mach number shocks \citep[e.g. Equations 4.1 and 4.2 of][]{Monaghan+1992}; we assume values of $\alpha = 0.25$ and $\beta = 0$ for the two artificial viscosity parameters following \citet{Ramachandran+2020}.

Finally, the equation of state describes the perturbations to the pressure from its initial value ($p_0 \approx 0$ at the fluid surface) due to adiabatic changes in the particle density \citep[e.g.][]{Cole+1948, Batchelor+2000,  Monaghan+2005}. That is,
\begin{equation}
p(\boldsymbol{r}) = \frac{\rho_0 {c_0}^{\!2}}{\gamma} \left[ \left( \frac{\rho(\boldsymbol{r})}{\rho_0} \right)^\gamma - 1\right]\,,
\end{equation}
where $\rho_0$ is the reference density, $c_0$ is the maximum sound speed expected in the system, and $\gamma = 7$ is an typically assumed adiabatic index for water. In this work, we set the density of water as $\rho_0 = 1000\rm\, kg\,m^{-3}$ and the sound speed as $c_0 = 32.85\rm\, m\,s^{-1}$; for computational efficiency, the reference sound speed in SPH simulations is typically chosen as ten times the largest velocity expected from to external forces to accurately capture the bulk motion of the fluid \citep{Monaghan+1992}. The sound speed of a given particle, as required in Equation \ref{tau}, is calculated from the reference sound speed and density as $c_s(\boldsymbol{r}) = c_0 [\rho(\boldsymbol{r})/\rho_0]^{(\gamma - 1)/2}$.

\subsubsection{Ice particle boundary conditions}

We model the ice as a rigid and impervious solid that is fixed in place throughout the simulation; i.e. we do not consider any elastic response in the ice or resonant coupling as in \cite{roeoesli2016seismic}. In the \emph{PySPH} framework, ice is thus represented by fixed-location particles that contribute to the acceleration of neighbouring fluid particles through the momentum equation. The density of the fluid particles increases as they approach the fixed-location ice particles, leading to an increased pressure which provides a repulsive force on the fluid particles. The repulsive force provided by the ice particles enforces the non-penetration boundary condition that $\boldsymbol{v} \cdot \boldsymbol{\hat{n}} = 0$ at the water-ice interface, where $\boldsymbol{\hat{n}}$ is a unit vector normal to the surface of the interface. The smoothed particle simulation requires a sizeable number of boundary particles to successfully enforce this non-penetration condition, often resulting in many more boundary than fluid particles in all but the highest resolution simulations (e.g. see Table \ref{tab:geometries}).

\begin{figure*}
\centerline{\includegraphics[width=0.8\textwidth,trim={65 15 85 45},clip]{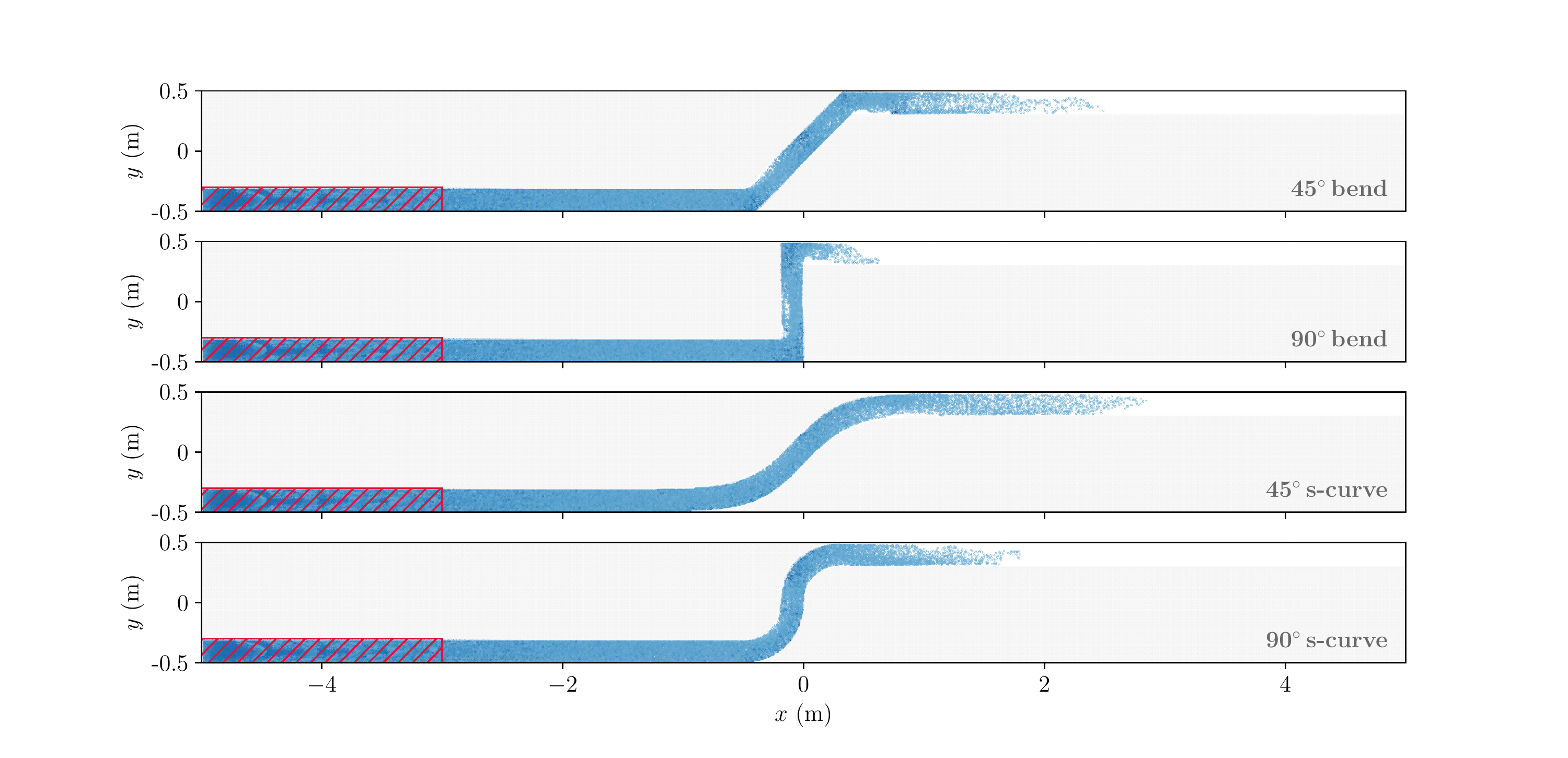}\vspace{4pt}}
\centerline{\includegraphics[width=0.8\textwidth,trim={65 15 85 45},clip]{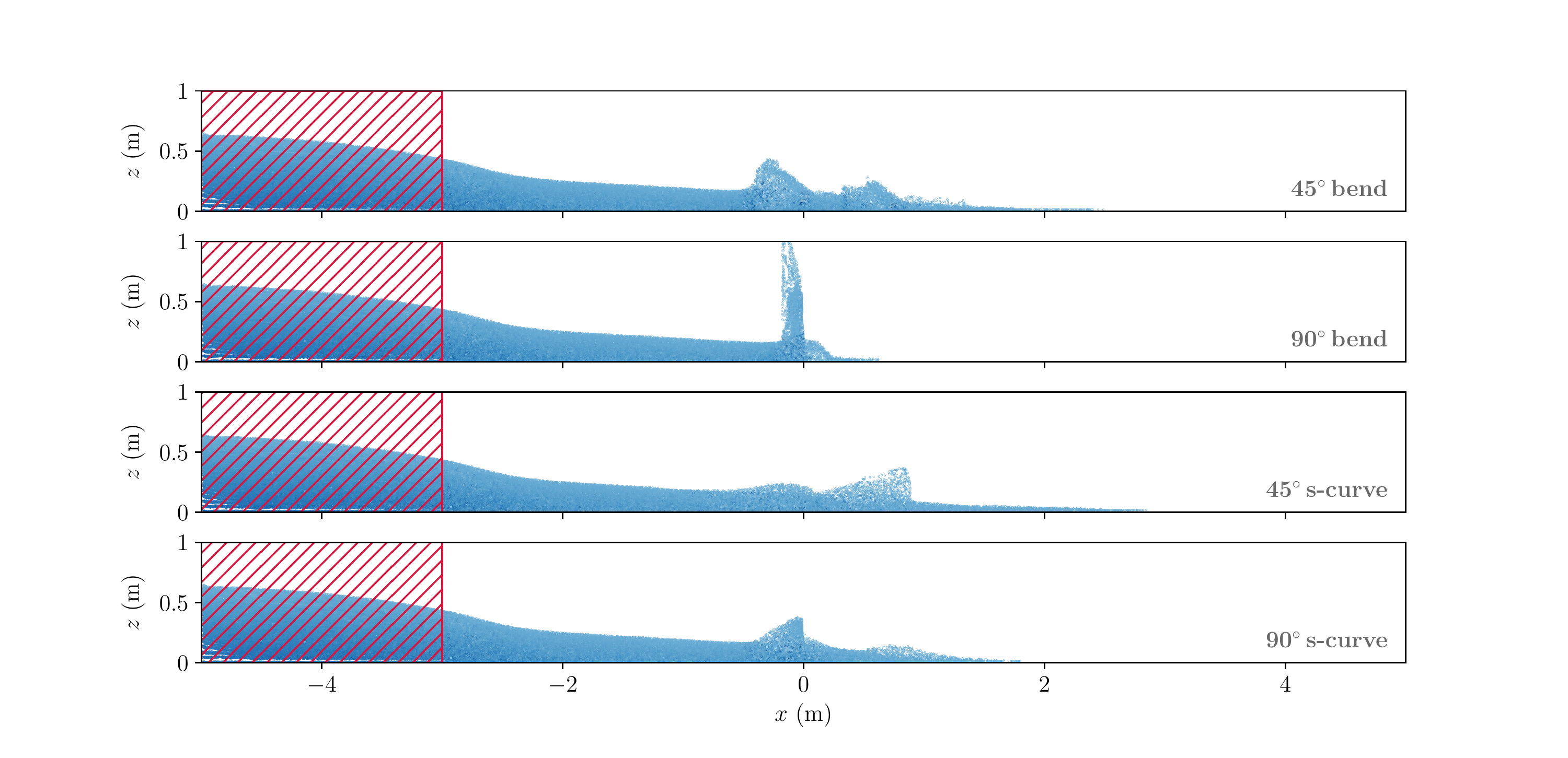}}
\caption{\textit{Top:} The four channel geometries viewed in the $xy$-plane (integrated along the $z$-direction). The water column (within the $0.2\rm\,m$ channel) is shown in blue at $t=1.5$\,s and the ice in pale grey; sections of the channel with no water are shown in white. The initial location of the water column at $t=0$ is shown by the hatched red rectangle. \textit{Bottom:} The four channel geometries viewed in the $xz$-plane (integrated along the $y$-direction); the ice is not shown for clarity. The modelling domain (equivalent to a 'tank' in a laboratory experiment) has a length $\Delta x = 10\rm\, m$ (edges are $x = \pm 5\rm\, m$), thickness of $\Delta y = 1\rm\, m$ (edges at $y = \pm 0.5\rm\, m$), and a height of $\Delta z = 2\rm\, m$ (edges at $z = 0$ and $z = 2\rm\, m$). The water body has an initial length $\delta x = 2\rm\, m$ (from $x = -5$ to $x = -3\rm\, m$), thickness $\delta y = 1\rm\, m$, and height $\delta z = 1\rm\, m$ (from $z = 0$ to $z = 1\rm\, m$).}
\label{fig:geometries}
\end{figure*}

\subsection{Fluid particle water-ice interface collisions}
\label{sec:Fluid particle water-ice interface collisions}

The SPH simulation calculates attributes of each fluid (and boundary) particle at regular time intervals, $t$. These include the Cartesian components position vector, $x$, $y$ and $z$, the Cartesian components of the velocity vector, $v_x$, $v_y$ and $v_z$, the particle mass, $m$, its density, $\rho$, and its pressure, $p$; the particle volume, $\delta V$, and cross-sectional area, $\delta A$, are derived from the mass and density. These attributes are output at a high temporal resolution, $\delta t = 0.001$\,s, to ensure the kinematics of each particle collision is captured in detail. We find that synthetic seismic signals converge to a common waveform when the temporal resolution is a factor of a few lower than the sample rate (i.e. $\delta t \lesssim 0.001$\,s is required for $200$\,Hz sampling).

In this work, these attributes and the known location of the water-ice interface are used to identify fluid particles that collide into the boundary. The fluid particle will interact with the water-ice interface if it passes within smoothing length $\epsilon$ of a boundary particle (i.e. the fluid particle kernel includes that boundary particle); in this work, we assume $\epsilon = h \delta l$ for resolution $\delta l$ and kernel scale $h \approx 2.6$\footnote{We present the smoothing length as the maximum extent of the kernel rather than the $1\sigma$ level.}. However, as the equilibrium fluid particle spacing is $\delta l = \epsilon/2.6$, low-energy interactions may be partially shielded by at least one other fluid particle closer to the boundary. Moreover, some equilibrium fluid particle locations may be approximately sited a smoothing length from a protruding edge or vertex along the boundary (e.g. $\epsilon \approx \sqrt{6}\delta l$\footnote{Possible closest distances to the boundary from equilibrium grid locations, assuming (e.g.) the boundary edges and vertices are all aligned with the grid, include (1, $\sqrt{2}$, $\sqrt{3}$, $2$, $\sqrt{5}$, $\sqrt{6}$, $\sqrt{8} $ and $3) \times \delta l$.}); such particles may frequently cross some critical distance from the boundary due to small amplitude oscillations about their equilibrium locations. In practice, to avoid including potentially millions of very low energy collisions, we require that fluid particles travel at least a distance of $0.2\epsilon$ towards the boundary beyond the smoothing length. This assumption will consequently exclude (or greatly reduce) the static pressure contribution to the force acting on the boundary. However, this is not relevant to the generation of seismic waves in this work, and as such, moderate changes to our chosen value (of $0.2\epsilon$) have no effect on our synthetic waveforms.

We track the path taken by the fluid particle whilst it is within a smoothing length $\epsilon$ of the boundary, and thus interacting with boundary particles at a potential collision site. The incident velocity, $\boldsymbol{v}_{0,n}$, is interpolated from the simulation outputs immediately before and after the $n$th fluid particle passes within a smoothing length of the boundary. Similarly, the reflected velocity vector, $\boldsymbol{v}_{\!\;\!\tau,n}$, is interpolated from the simulation outputs immediately before and after the $n$th fluid particle exits a region of smoothing length $\epsilon$ from any boundary particle.
The particle exchanges its kinetic energy into elastic potential energy as it approaches closer to the boundary than these two locations, with the minimum kinetic energy reached at the stationary point in its trajectory. However, we cannot directly measure the location of this stationary point due to the high particle velocities (up to $|v| = 5\rm\, m\,s^{-1}$) relative to the temporal resolution of the simulation outputs ($\delta t = 0.001$\,s); i.e. the particle position may only be sampled every 0.005\,m compared to the smoothing length of 0.026\,m (resolution discussed in Section \ref{sec:Water-ice simulation geometry}). We therefore assume the mid-point of the particle trajectory within a smoothing length of the boundary is representative of the stationary point. The time of the collision, $t_n$, is derived from two closest simulation outputs to this mid-point path length, whilst the residual velocity vector $\boldsymbol{v}_{\!\;\!\star,n}$ (see Section \ref{sec:Collision as a damped harmonic oscillator}) is estimated by interpolating the Cartesian components of the velocity from these two outputs.

The SPH simulation prevents fluid particles from directly impacting the water-ice interface due to the finite width of the kernel, and, as a result, the stationary point on a given particle trajectory will not lie on the boundary. We therefore assume the location on the interface closest to the stationary point of the $n$th particle trajectory is the actual collision site, $\boldsymbol{r}_{\!\;\!n}$. This is achieved by numerically approximating the unit normal to the surface in the region of the expected collision site through the fitting of a plane to the locations of the proximate boundary particles. This permits an estimate of the location of the collision site to within the precision of the simulation resolution. The orientation of the interface (as described by the unit normal) is additionally required to ensure the synthetic seismic waves are only propagated through the ice, not back into the water. 
The fluid particles involved in collisions with the water-ice interface, and the properties derived for their interactions from the simulation outputs, can now be used to generate synthetic seismic waves based on our analytic model.

\subsection{Water-ice simulation geometry}
\label{sec:Water-ice simulation geometry}

We model the flow of water along narrow channels of different geometries set several metres into the ice, aiming to capture the general characteristics of the seismic signals resulting from the variety of supraglacial flows, and provide demonstration cases for the signals thus generated. The models are based on the well-studied dam-break problem in which a body of water in one part of a tank collapses under gravity and impacts a wall at the other end of the tank. 

The water flows in the positive $x$-direction along channels of constant thickness ($0.2\rm\,m$) with ice particles placed within the modelling domain to prescribe different paths in the $xy$-plane for each geometry. The four channel geometries (Figure \ref{fig:geometries}) are: (i) two 45$^{\circ}$ bends, to the left then to the right; (ii) two 90$^{\circ}$ bends; (iii) a gentle 45$^{\circ}$ s-curve, described using a hyperbolic tangent function; and (iv) a sharp 90$^{\circ}$ s-curve, described using two circular arcs. The origin of our coordinate system is located at the centre of the two bends. The number of fluid and ice particles in the four simulations is summarised in Table \ref{tab:geometries}. The volume of each fluid particle is initially $\delta V = 1\,\rm cm^3$, corresponding to a spatial resolution of $\delta l = 0.01$\,m; this is a trade-off between computational efficiency and ensuring the spatial resolution is sufficiently high such that the collision duration (derived from the impulse; see Section \ref{sec:Droplet collisions with the water-ice interface}) is shorter than the sample rate.

\begin{table}
\begin{center}
\caption[]{Number of ice and fluid (water) particles in the smoothed particle simulations for the four supraglacial channel geometries.}
\label{tab:geometries}
\renewcommand{\arraystretch}{1.1}
\setlength{\tabcolsep}{6pt}
\begin{tabular}{cccc}
\hline\hline
Geometry&Fluid particles&Ice particles&Total particles \\
\hline
45$^\circ$ bend&380\,000&12\,937\,176&13\,171\,176 \\
90$^\circ$ bend&380\,000&12\,803\,526&13\,183\,526 \\
45$^\circ$ s-curve&380\,000&13\,065\,126&13\,445\,126 \\
90$^\circ$ s-curve&380\,000&13\,044\,576&13\,424\,576 \\
\hline
\end{tabular}
\end{center}
\end{table}

\section{Results}
\label{sec:results}

We apply the analytic method developed for the generation and propagation of synthetic seismic waves (Section \ref{sec:Seismic wave generation and propagation}) to the outputs of the four SPH simulations (Section \ref{sec:Smoothed particle hydrodynamic simulations}) corresponding to the different channel geometries (Figure \ref{fig:geometries}). We qualitatively examine the characteristics of these synthetic seismic signals to appraise our methodology and to gain further insight into the generated waveforms. In particular, we investigate the seismic wavefield detected at the surface within an $\sim$\,$1$\,km radius of each channel (Section \ref{sec:Surface seismic wavefield}), the synthetic time series and spectral density at a single seismometer location (Section \ref{sec:Synthetic time series}), and the relative strengths of the transverse and longitudinal waves as a function of angle around the channel centre (Section \ref{sec:Directional dependence of seismic wavefield}).

\subsection{Seismic wavefield at the ice surface}

\begin{figure*}
\centerline{\includegraphics[width=0.95\textwidth,trim={70 38.75 45 38.75},clip]{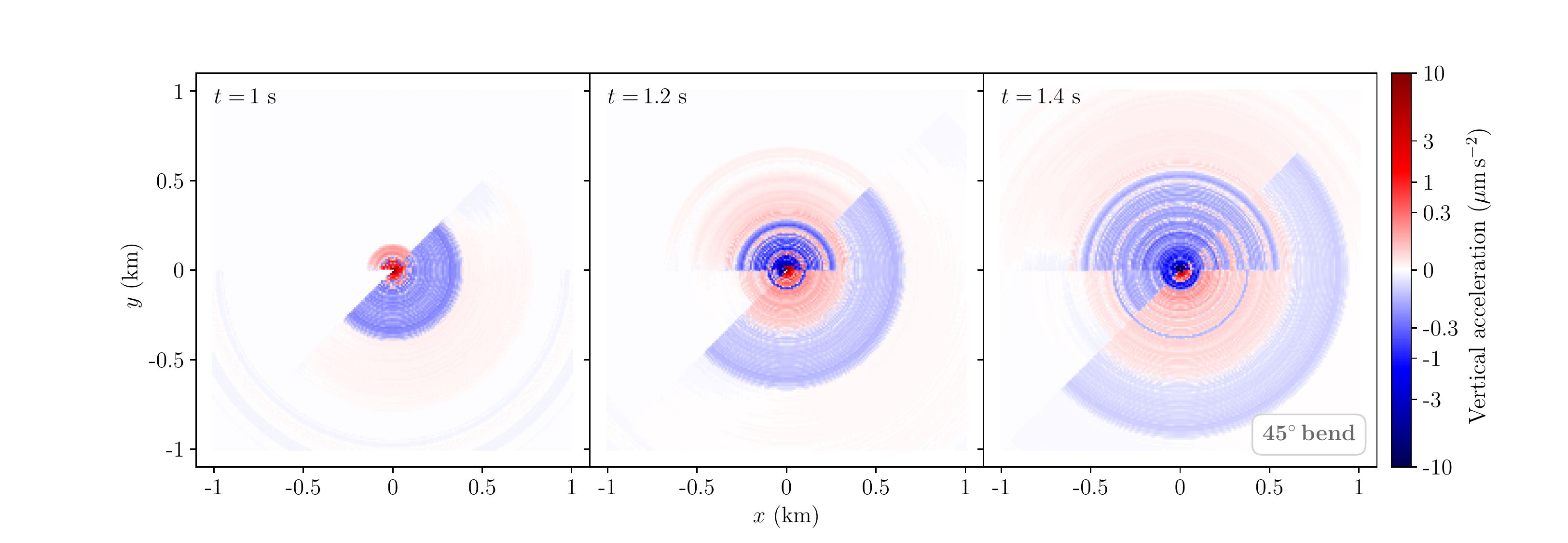}\vspace{3pt}}
\centerline{\includegraphics[width=0.95\textwidth,trim={70 38.75 45 38.75},clip]{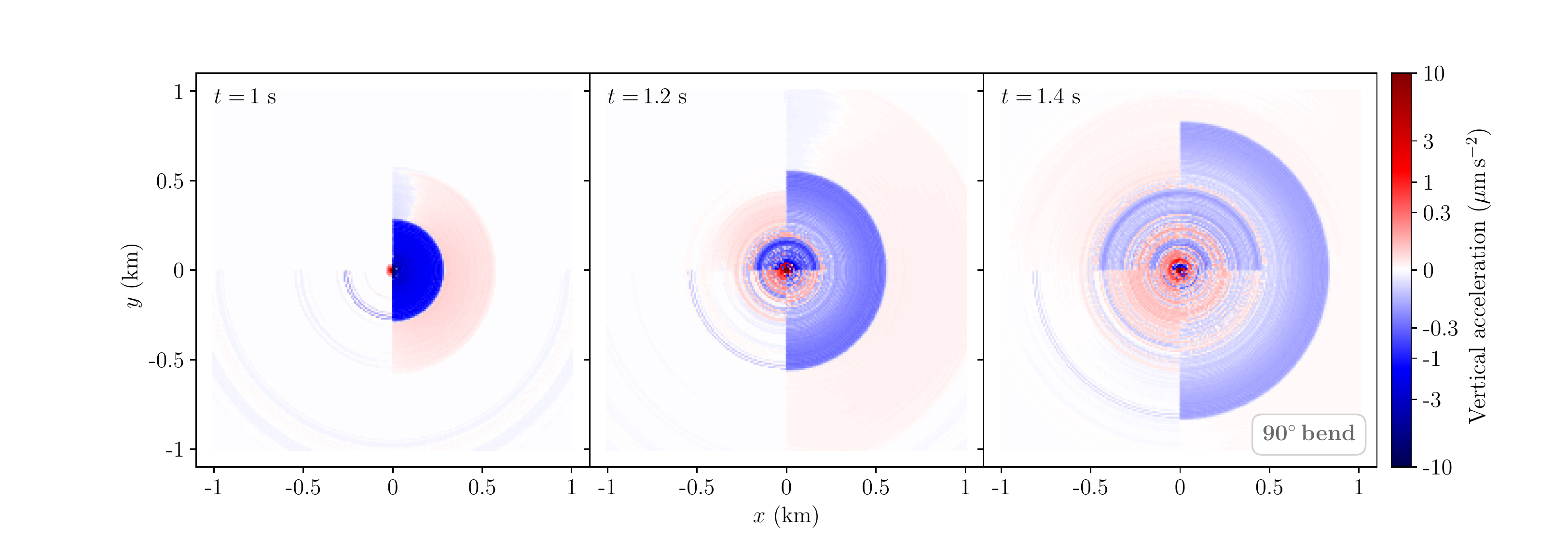}\vspace{3pt}}
\centerline{\includegraphics[width=0.95\textwidth,trim={70 38.75 45 38.75},clip]{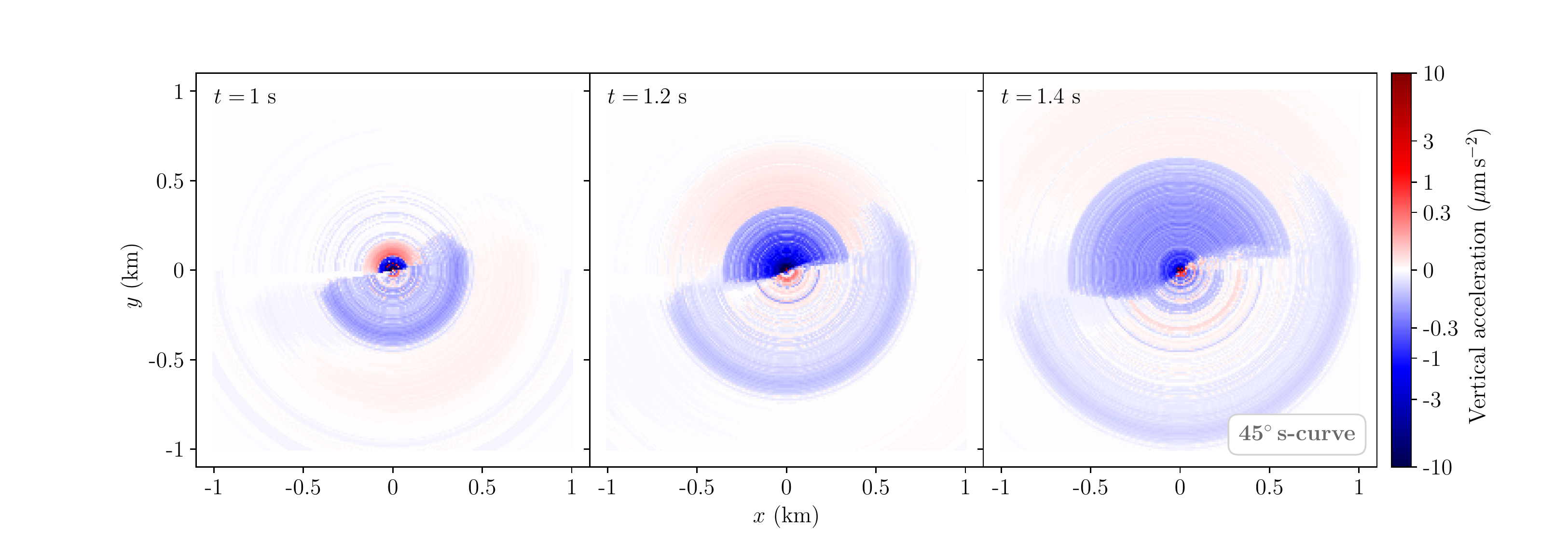}\vspace{3pt}}
\centerline{\includegraphics[width=0.95\textwidth,trim={70 7.5 45 38.75},clip]{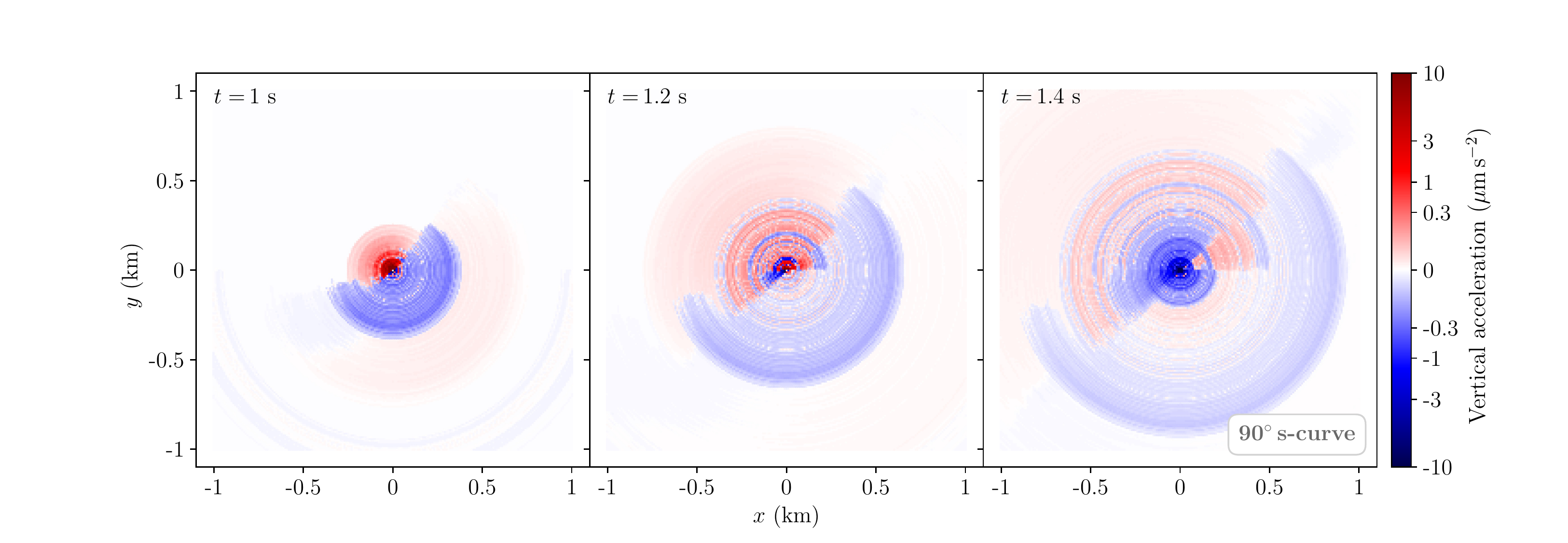}}
\caption{Seismic wavefield simulated at grid locations across a flat surface ($xy$-plane) 2\,m above the given channel, with flow in the positive $x$-direction. The $z$-component (vertical) of the acceleration is shown, with red corresponding to positive accelerations, blue to negative accelerations, and white to locations with no signal. The colourbar is stretched as $\text{sgn}\, a \sqrt{|a|}$ so that values at greater distance from the collision can be seen (equivalent to a reduced $1/r$ radiation pattern geometry). The seismic wavefield is shown for snapshots at three times after the collision at the first bend ($t \approx 0.75$ s): $t=1$, $t=1.2$, and $t=1.4$\,s. \textit{Top row:} 45$^\circ$ bend  \textit{second row:} 90$^\circ$ bend, \textit{third row:} 45$^\circ$ s-curve, and \textit{bottom row:} 90$^\circ$ s-curve.}
\label{fig:surface}
\end{figure*}

We capture signals across an ice surface sited 2\,m above the base of the channels, sampling at 10\,m intervals ($x$- and $y$-directions) on a Cartesian grid out to an $\sim$\,$1$\,km radius from the channels (i.e. $x=\pm1$\,km and $y=\pm1$\,km). We generate synthetic acceleration waveforms as a function of time at these locations and present the vertical component of the signal (Figure \ref{fig:surface}).

\label{sec:Surface seismic wavefield}

The primary water-ice collision in the 45$^\circ$ bend channel occurs against the first bend (sited at $x \approx -0.3$ and $y \approx -0.4$\,m; see Figure \ref{fig:geometries}) at $t \approx 0.7$\,s. We show the resulting surface seismic wavefield at three later times in Figure \ref{fig:surface} (top row). The seismic waves generated by this first collision (i.e. those seen at $t=1$\,s) radiate outwards in a semi-circular pattern towards $(x,y) = (1,-1)$\,km, restricted to azimuthal angles from $\phi = -135^\circ$ to $+45^\circ$ (i.e. the ice-side of the interface). The longitudinal wave amplitude is greatest along an azimuthal angle of $\phi \approx -45^\circ$, however, the total wave amplitude is only marginally weaker off-axis due to the high assumed transverse-to-longitudinal wave amplitude ratio of $\Lambda = 0.5$. 
The short duration of the primary collision against the interface presented by first of the two bends results in a narrow wave packet of approximately 300\,m in width; at later times this leads to a pronounced arc of high-amplitude vertical accelerations. The secondary water-ice collision against the second bend in the channel at $t \approx 1$\,s (sited at $x\approx 0.5$ and $y\approx0.5$\,m) generates another semi-circular arc of seismic waves; the seismic waves radiate along the positive y-axis and are bounded by azimuthal angles from $\phi = 0$ to $+180^\circ$. The arrival of smaller amplitude signals due to the turbulence of the reflected flow from these two collisions leads to lower amplitude features seen from $t=1.4$\,s.

The seismic wavefield patterns generated by the collisions for the other channel geometries are qualitatively the same as the 45$^\circ$ bend channel, however, their timing, strength and orientation vary greatly (Figure \ref{fig:surface}, lower three rows). The semi-circular arcs correspond to the azimuthal angles on the ice-side of the interface at locations with energetic particle collisions. The peak wave amplitude (of the primary or secondary collision) is approximately centred on these semi-circular arcs, and we investigate the angular dependence of the longitudinal (P) and transverse (S) waves in detail later in this section.

\subsection{Synthetic time series}
\label{sec:Synthetic time series}

We discuss the synthetic waveforms predicted for a seismometer located at $(x, y, z) = (10, 5, 2)$\,m; i.e. 10\,m downstream of the centre of the two bends on the channel, shifted 5\,m in the horizontal normal to the direction of flow, and raised 2\,m above the base of the channel. This location is 11.36\,m from the primary collision site at an azimuthal angle of $\phi = 26.6^\circ$ (from positive $x$- towards positive $y$-axis) and an inclination angle of $\theta = 11.3^\circ$ (from positive $x$- towards positive $z$-axis). This off-axis location is chosen to ensure the longitudinal (P) and transverse (S) wave components are not aligned with the components ($x,y,z$) at the seismometer.

%\subsubsection{Three-component acceleration time series}

\begin{figure*}
\centerline{\includegraphics[width=0.9\textwidth,trim={57.5 35 85 35},clip]{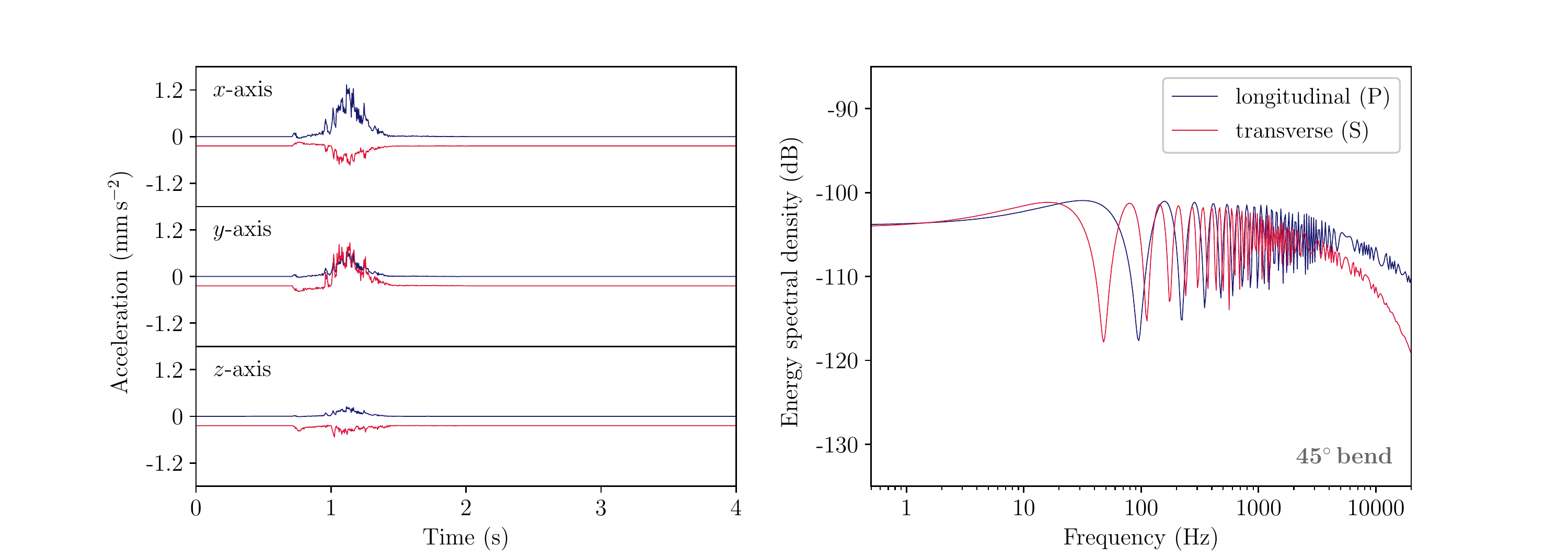}\vspace{6pt}}
\centerline{\includegraphics[width=0.9\textwidth,trim={57.5 35 85 35},clip]{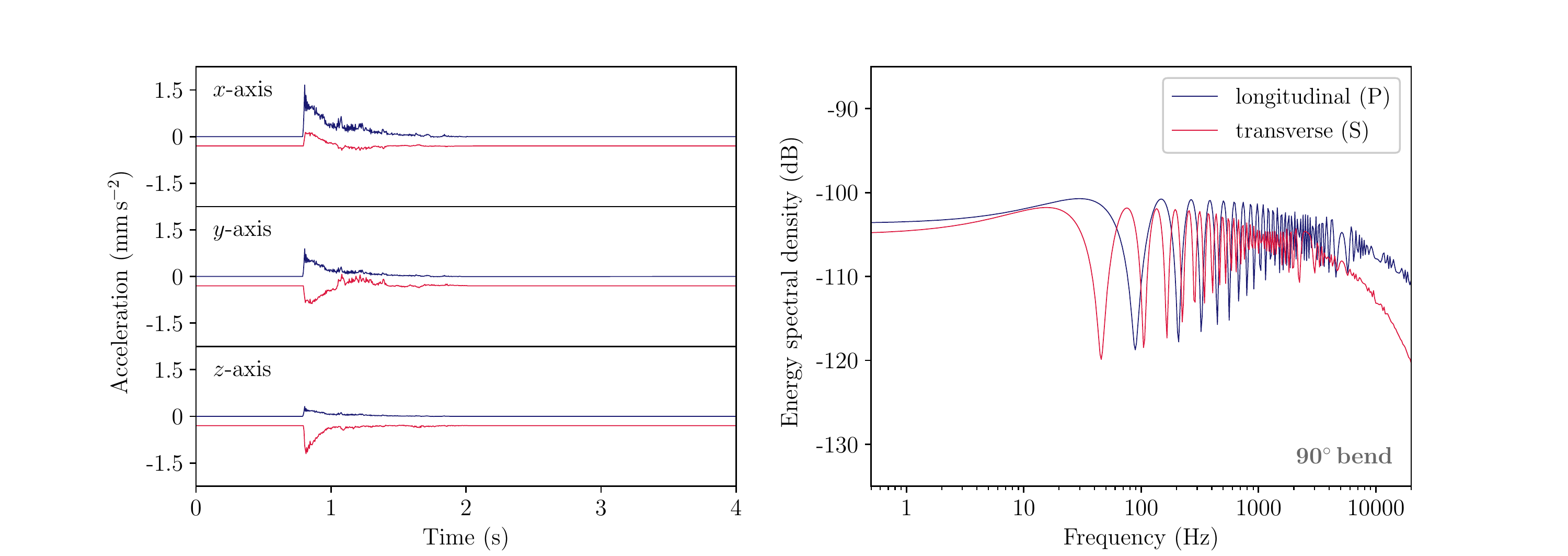}\vspace{6pt}}
\centerline{\includegraphics[width=0.9\textwidth,trim={57.5 35 85 35},clip]{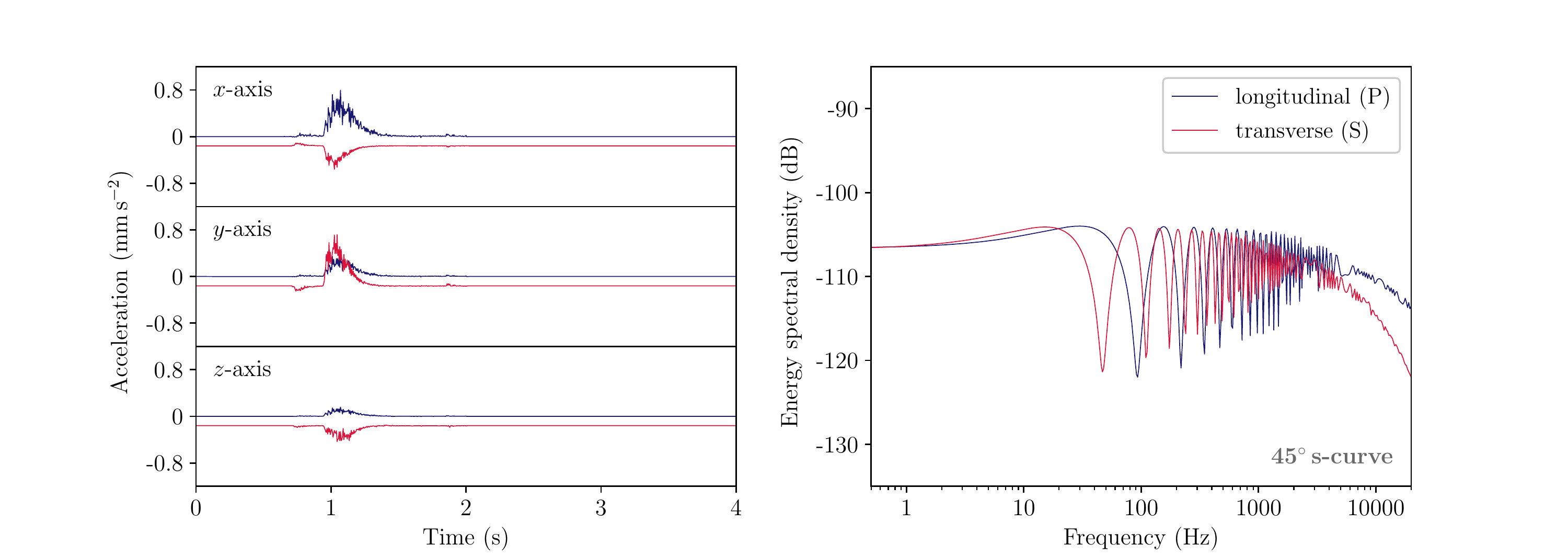}\vspace{6pt}}
\centerline{\includegraphics[width=0.9\textwidth,trim={57.5 2.5 85 35},clip]{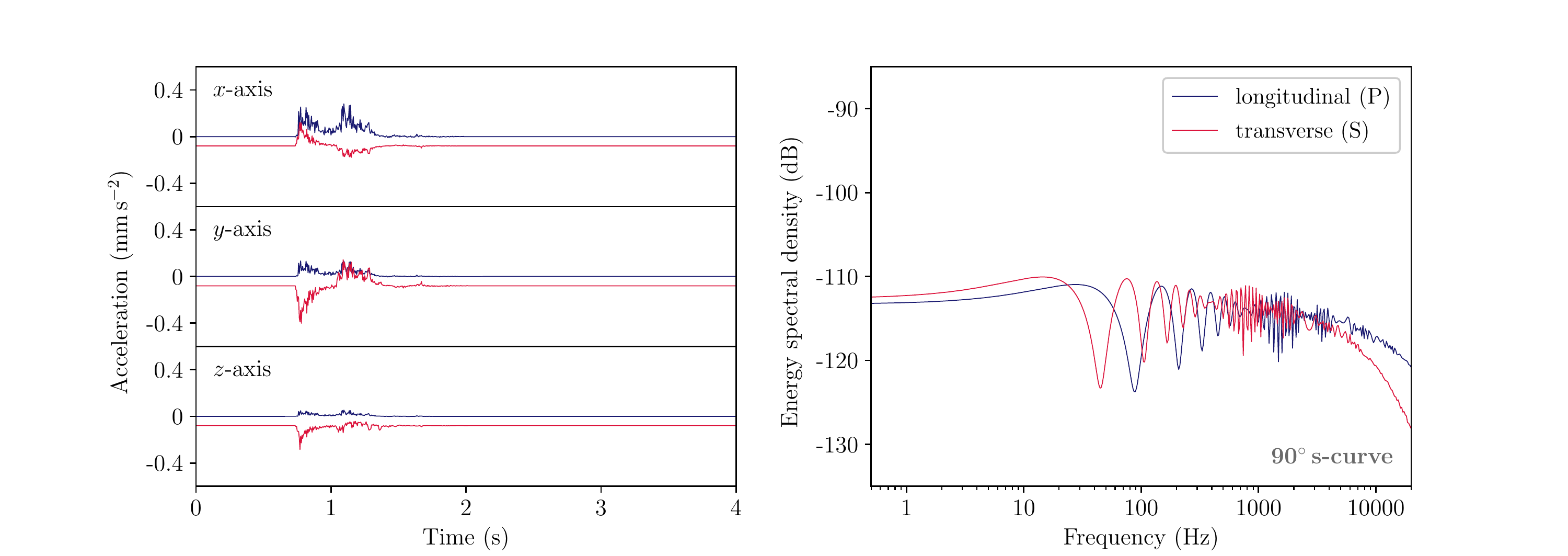}}
\caption{\textit{Left column:} Synthetic seismic wave acceleration amplitude at a seismometer located at $(x, y, z) = (10, 5, 2)$\,m as a function of time, $t=0$ being the start of the simulation. The $x$, $y$ and $z$ components of the signal are shown for longitudinal (P, blue) and transverse (S, red) waves, the latter offset by $-20\%$ the peak amplitude for clarity. The hydrodynamic simulation is only run out to $t=2$\,s, but we present a longer time series for consistency with later plots. \textit{Right column:} Energy spectral density for the same collision including all three Cartesian components of the signal. The channel geometries are:\textit{top row:} $45^\circ$ bend, \textit{second row:} 90$^\circ$ bend, \textit{third row:} 45$^\circ$ s-curve, and \textit{bottom row:} 90$^\circ$ s-curve.}
\label{fig:waveform}
\end{figure*}

Synthetic three-component acceleration waveform data generated for the 45$^\circ$ bend channel geometry is shown in Figure \ref{fig:waveform} (top-left; with separated longitudinal and transverse components). The first arrival of seismic waves occurs at $t \approx 0.7$\,s, resulting from the water-ice collision against the first of the two bends in the channel (primary collision). The fluid particles are deflected at the first bend without losing significant energy (due to the shallow angle of incidence) leading to only a weak detection. However, the flow continues to rapidly accelerate towards the second bend due to pressure from the collapsing water column. As a result, the strongest seismic signals for this channel geometry are associated with the second bend (secondary collision) peaking at $t \approx 1.1$\,s.
The peak longitudinal (P) wave amplitude is directed at an azimuthal angle of $\phi \approx 85^\circ$ (see Section \ref{sec:Directional dependence of seismic wavefield}), and thus only a (relatively) moderate acceleration amplitude is detected at the chosen seismometer location ($\phi = 26.6^\circ$; i.e. $\approx 60^\circ$ from the peak), primarily in the $x$-component. By contrast, the peak transverse (S) wave amplitude is directed at pairs of azimuthal and inclination angles normal to the peak P wave amplitude; i.e. azimuthal angles of $\phi \approx 25^\circ$ and $155^\circ$ for the inclination angle of the seismometer, the former of which is very close to the seismometer location. As a result, the secondary collision is (relatively) strongly detected through the S wave arrivals in the $y$- and $z$-components of the seismometer.

The synthetic acceleration waveform data for the 90$^\circ$ bend channel geometry has an entirely different character (Figure \ref{fig:waveform}, second row-left) due to the differing strengths of the primary and secondary collisions. The fluid particles collide normal to the water-ice interface at the first bend losing the majority of their energy in a highly energetic interaction. However, these fluid particle consequently have low reflected velocities and tend to `shield' the interface from further high-energy particle collisions (i.e. the reflected particles cannot get out of the way of newly incident particles). 
The secondary collision is subsequently also relatively muted due to the low reflected velocities and turbulent nature of the flow (Figure \ref{fig:geometries}). For this channel geometry, the (primary) P wave arrival is strongly detected by the $x$-component of the seismometer; the peak P wave amplitude is directed at an azimuthal angle of $\phi \approx -10^\circ$ with a sizeable acceleration amplitude detectable at the seismometer (i.e. $\approx 35^\circ$ from the peak). The peak S wave amplitude is similarly proximal to the seismometer location and, consequently, the S wave arrival from the secondary collision is also strong.

The relative amplitudes of the acceleration provided to the three seismometer components (for any of the channel geometries) additionally provides a means of appraising our method. The seismometer is located at $(x, y, z) = (10, 5, 2)$\,m and therefore, considering a source at the origin (approximately true for all water-ice collisions), the P wave acceleration along the $x$-component of the seismometer will be double that detected by the $y$-component, and five-fold greater than that detected by the $z$-component of the seismometer. This qualitative expectation is consistent with our model predictions (e.g.) in Figure \ref{fig:waveform} (second row-left). Meanwhile, the S wave acceleration should primarily be detected in at least one of the components of the seismometer normal to the direction of propagation; in this case, the S wave is most strongly detected in both the $y$- and $z$-components with only a modest contribution seen in both the $x$-component.

The synthetic three-component signals for the other channel geometries are shown in the lowest two panels (Figure \ref{fig:waveform}), again with impact times and other characteristics logical for the geometry of the smoothed particle hydrodynamic simulation. The collision of the water column with the first bend in each of the channels is again detectable at $t \approx 0.7$\,s, albeit with a very low amplitude for the gentle 45$^\circ$ s-curve. The primary collision in the 45$^\circ$ s-curve channel is associated with very shallow angles of incidence leading to minimal energy loss during the particle collisions. As a result, the flow accelerates towards the second bend whilst impacting the second bend at a marginally steeper angle (due to the angle of reflectance from the first band) leading to higher amplitude signals. 
By contrast, the 90$^\circ$ s-curve geometry has a comparatively sizable energy loss at the first bend and thus a (relatively) strong seismic signal; this occurs as  sections of the water-ice interface present to the incoming flow at angles of incidence up to approximately 45 degrees (e.g. at $x\approx-0.1$ and $y\approx-0.3$\,m). However, unlike for the standard 90$^\circ$ bend channel, the flow remains somewhat laminar after this collision and thus can continue to accelerate towards the second bend leading to a secondary peak in the seismic waveform.

\subsection{Spectral density}

We now consider the spectral properties of seismic signals owing to the fluid particle collisions on the water-ice interface. Neither the energy nor the power spectral density is invariant to the assumed collision duration. By contrast, the acceleration-squared spectral density is invariant to both the collision duration and the simulation resolution, whilst differing from the energy and power spectral densities by an approximately constant factor. We therefore derive an expression for the acceleration-squared spectral density based on the Fourier transform of the time-varying force at the seismometer location, $\mathcal{F}_{n}(\boldsymbol{r}, \xi)$ (given in Equation \ref{force_transform}). That is,
\begin{equation}
S(\boldsymbol{r}, \xi) = \frac{1}{(\varrho \delta V)^2} |\mathcal{F}_{n}(\boldsymbol{r}, \xi)|^2 \,,
\end{equation}
where the longitudinal and transverse components of the applied force need to be considered separately due to their different weak dispersion and attenuation relations. The acceleration-squared spectral density is presented using the decibel scale in this work, with a reference spectral density of $1\,\rm (m\, s^{-1})^2\, Hz^{-1}$, to equally represent the energy and power spectral density; as a result, we hereafter refer to this quantity as the energy spectral density.

The energy spectral density of seismic waveforms detected by a seismometer located at $(x, y, z) = (10, 5, 2)$\,m is shown in the right panels of Figure \ref{fig:waveform} for the four channel geometries. The total energy carried by the seismic wave to the seismometer is equivalent for each of the channel geometries, albeit the energy received from the 90$^\circ$ s-curve is significantly lower. This is consistent with the overall lower acceleration amplitudes provided to the seismometer for this channel geometry. 

The longitudinal (P) wave carries either comparable or more energy than the transverse (S) wave, as expected for the assumed transverse-to-longitudinal wave amplitude ratio of $\Lambda = 0.5$. However, the fraction of energy carried by the P and S waves is a strong function of frequency: the energy carried differs by less than a factor of two at frequencies $\xi < 100$\,Hz, but beyond $\xi > 10$\,kHz the S waves carry only one-tenth of the total wave energy. This results from the highly frequency-dependent energy losses (i.e. attenuation) of the anelastic ice (Section \ref{sec:Seismic waves}) which becomes more significant with distance from the collision site. We discuss such changes to the energy spectral density, and consequently the shape of the wave packet, in Section \ref{sec:discussion}.

\subsection{Directional dependence of seismic wavefield}
\label{sec:Directional dependence of seismic wavefield}

The directional dependence of the seismic wave induced acceleration is investigated by generating synthetic signals for seismometer locations in a sphere of radius 10\,m surrounding each supraglacial channel geometry. We calculate the time-averaged longitudinal (P) and transverse (S) wave acceleration at azimuthal angles from $\phi = -180^\circ$ to 180$^\circ$ (i.e. from positive $x$- towards positive $y$-axis), and inclination angles from $\theta = -90^\circ$ to 90$^\circ$ (i.e. from positive $x$- towards positive $z$-axis); the grid spacing in both azimuth and inclination is set at 5 degrees. The time-averaged acceleration is calculated considering the magnitude of the three-component acceleration time series over a duration of one second from the first arrival at $t \approx 0.7$\,s (this time is assumed for all angles).

\begin{figure*}
\centerline{\includegraphics[width=0.725\textwidth,trim={125 33.75 117.5 35},clip]{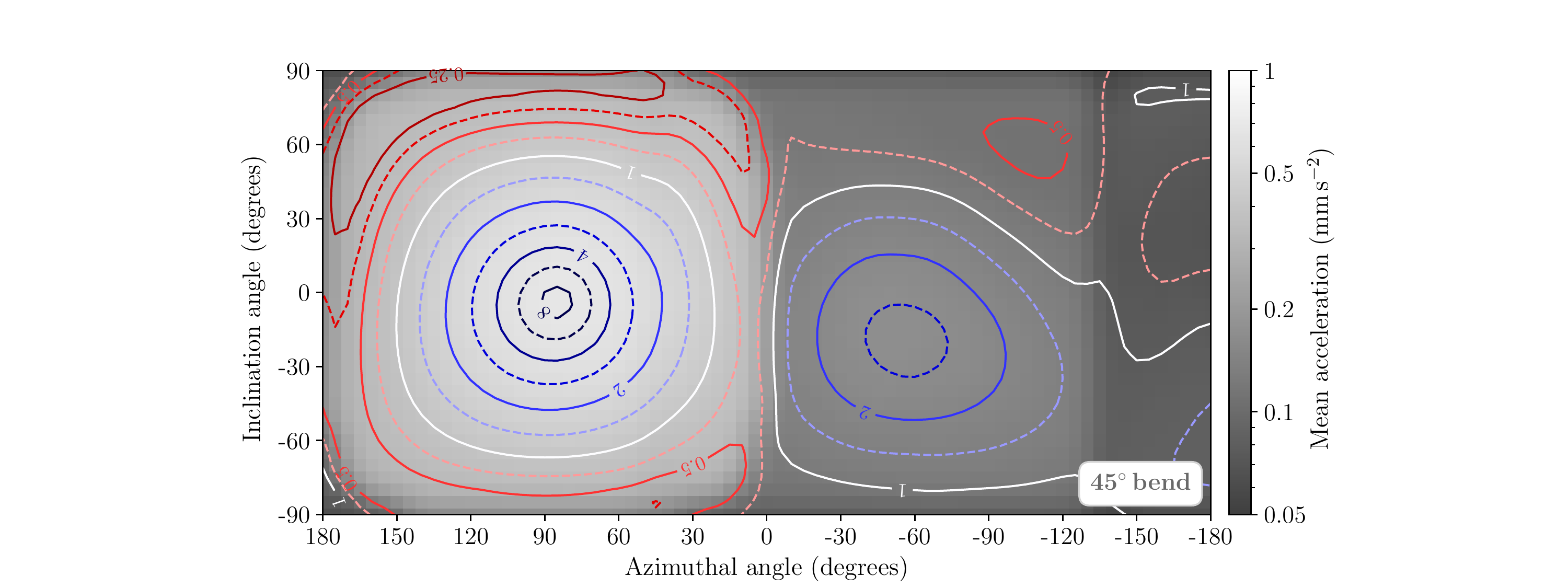}\vspace{4pt}}
\centerline{\includegraphics[width=0.725\textwidth,trim={125 33.75 117.5 35},clip]{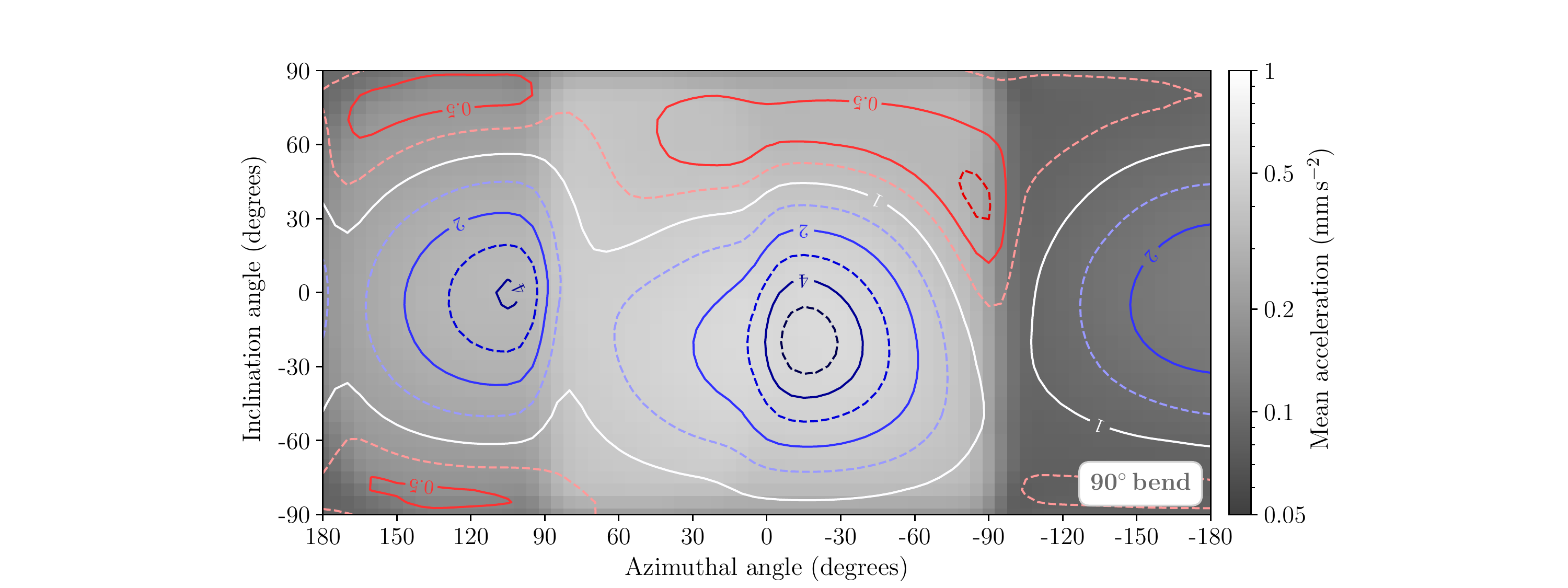}\vspace{4pt}}
\centerline{\includegraphics[width=0.725\textwidth,trim={125 33.75 117.5 35},clip]{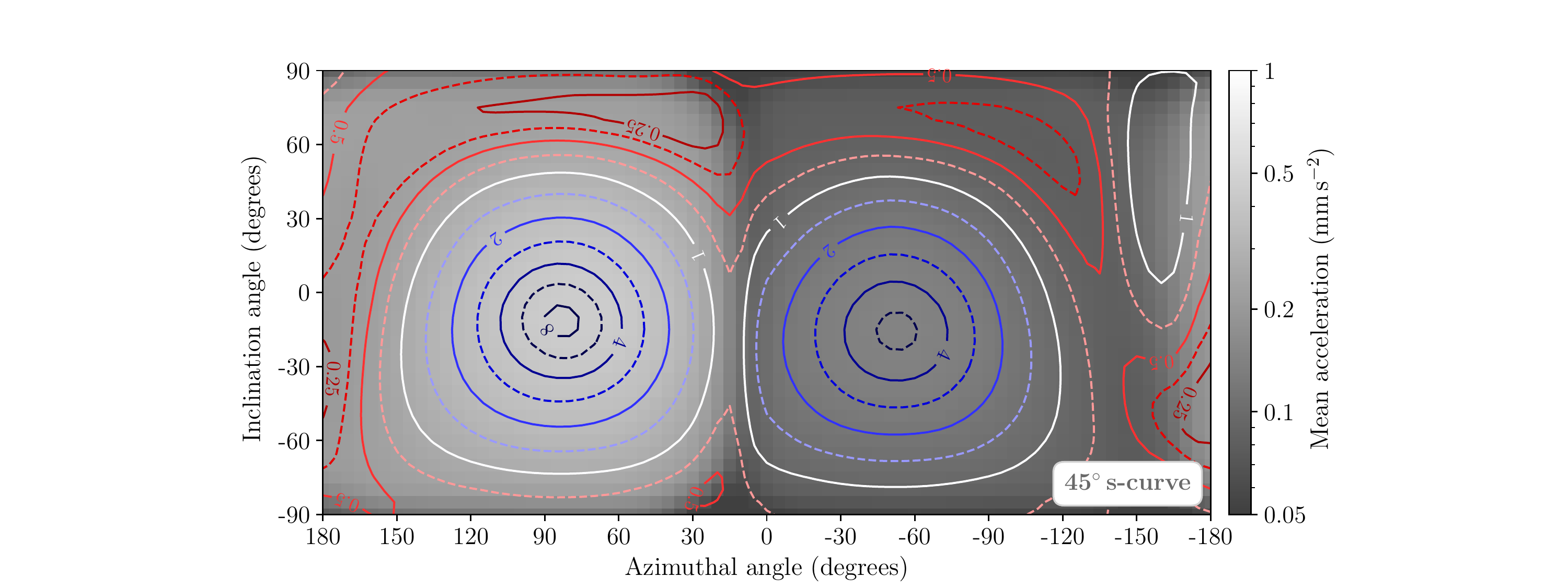}\vspace{4pt}}
\centerline{\includegraphics[width=0.725\textwidth,trim={125 2.5 117.5 35},clip]{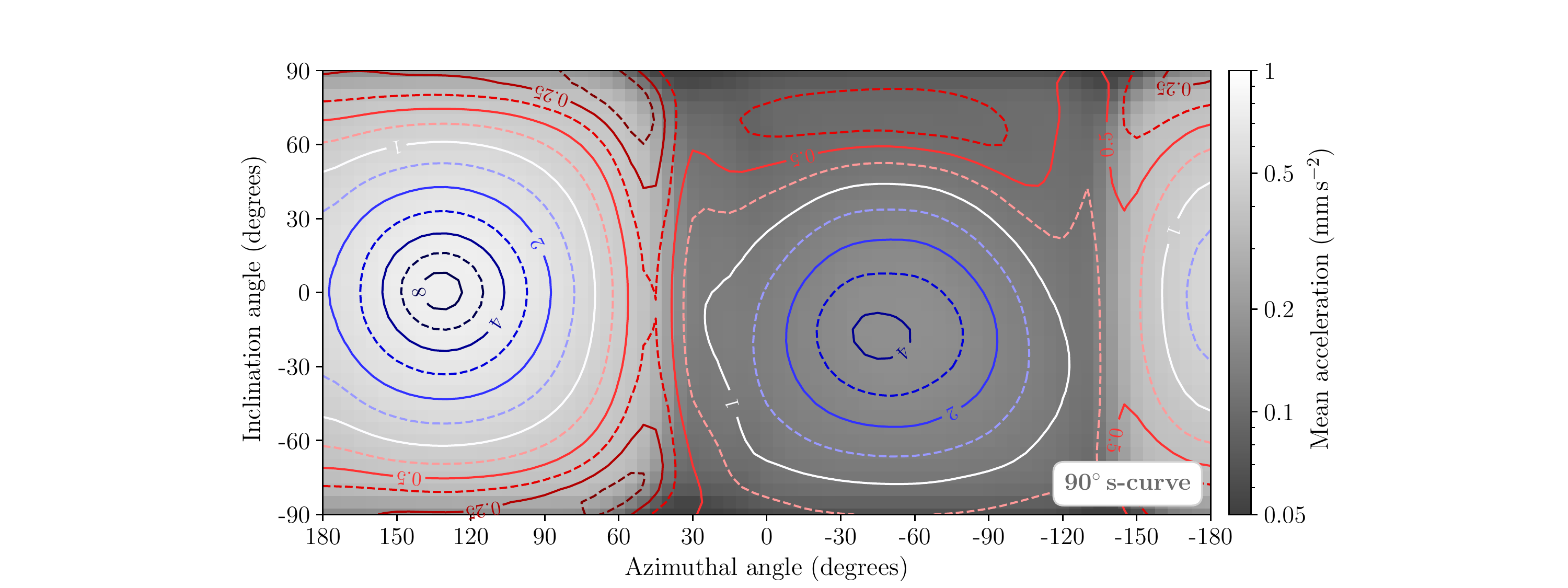}}
\caption{Time-averaged magnitude of the acceleration for seismic waves detected 10\,m from the centre of the two bends as a function of the azimuthal angle, $\phi$ (from positive $x$- towards positive $y$-axis), and inclination angle, $\theta$ (from positive $x$- towards positive $z$-axis). The shading shows the mean (time-averaged) magnitude of the acceleration measured as the quadrature sum of the $x$, $y$ and $z$ components of the signal at the seismometer. The ratio of the mean longitudinal (P) to mean transverse (S) acceleration is shown by the contours, with the ratios annotated on the contours for each factor of two. The channel geometries are: \textit{top row:} 45$^\circ$ bend, \textit{second row:} 90$^\circ$ bend, \textit{third row:} 45$^\circ$ s-curve, and \textit{bottom row:} 90$^\circ$ s-curve.}
\label{fig:direction}
\end{figure*}

The time-averaged magnitude of the acceleration for seismic waves generated by water flowing in the 45$^\circ$ bend channel is shown in Figure \ref{fig:direction} (top) as a function of the seismometer location. Two local maxima in the P wave acceleration, associated with the primary and secondary collisions, occur at $(\phi, \theta) = (-15^\circ, -50^\circ)$ and $(0, 85^\circ)$ with amplitudes of $0.03\rm\, mm\, s^{-2}$ and $0.13\rm\, mm\, s^{-2}$ respectively. The half plane associated with the more energetic secondary collision consequently has the highest total accelerations, though with amplitudes slowly reducing away from the P wave peak as an increasing fraction of the seismic wave energy is carried by the weaker transverse waves. The S wave amplitude peaks in a circular ring an angular distance 90$^\circ$ from each of the two P wave peaks as a result of their orthogonal generation mechanisms; the rings from the primary and secondary collisions coincide in a horizontal band centred at approximately $(70^\circ, 50^\circ)$ leading to a region of high S wave accelerations ($0.06\rm\, mm\, s^{-2}$) when averaging over both collisions.

The angular dependence of the time-averaged acceleration produced by the collisions in the other three channel geometries is shown in the bottom three panels of Figure \ref{fig:direction}. The 90$^\circ$ bend channel has qualitatively the same pattern as for the 45$^\circ$ bend, albeit the P wave peak for the primary collision ($0.10\rm\, mm\, s^{-2}$) is this time stronger than that of the secondary collision ($0.06\rm\, mm\, s^{-2}$). The location of both peaks are shifted `leftwards' to $(\phi, \theta) = (-10^\circ, -20^\circ)$ and $(0, 105^\circ)$ respectively, somewhat closer in separation than for the 45$^\circ$ bend geometry. This results in a vertical band, from approximately $\theta = 0$ to $90^\circ$, with the highest total accelerations (white vertical stripe in Figure \ref{fig:direction}, top).
Meanwhile, the location and relative amplitudes of the P wave peaks for the two s-curve channels are broadly consistent with the 45$^\circ$ bend geometry. Notably, the seismic wavefield for the 90$^\circ$ s-curve channel does not at all resemble that of the 90$^\circ$ bend geometry. The laminar flow in the 90$^\circ$ s-curve channel leads to time-averaged accelerations (at the second bend) that are nearly double that of the more turbulent 90$^\circ$ bend (at its first bend).

\section{Discussion}
\label{sec:discussion}

\subsection{Limitations}
\label{sec:Limitations}

The impulse capturing method presented in this work to track fluid particle collisions may encounter difficulties characterising some flow types due to our requirement that particles traverse some nominal distance within a set critical length-scale of the boundary (see Section \ref{sec:Fluid particle water-ice interface collisions}); this assumption greatly improves the computational efficiency of the algorithm by filtering out low-energy particle oscillations (about their equilibrium position) with amplitudes much less than $\delta l$. This technique is therefore of less utility if considering less energetic fluid motions such as pressure waves in stationary water, or particles colliding at very shallow angles or moving parallel to the ice as could be the case with turbulent flow. Further, in some energetic flows (e.g. $90^\circ$ bend channel), fluid particles with low reflected velocities may `shield' the interface from further high-energy particle collisions; i.e. the shielding particles will oscillate in response to the collision transferring the energy to the boundary as a pressure wave. Such situations may be handled, to some extent, by introducing numerous finely-spaced critical length-scales (e.g. over the range $2\delta l$ to $3\delta l$) to capture small amplitude oscillations equally-well from any equilibrium particle location, and thus abandon the need for a low-energy filter to prevent collisions being non-uniformly selected across the particle grid. 

This would enable very low energy interactions between particles to be captured, including those due to static pressure, but at the cost of adding millions of collisions to the seismic wave calculation (compared to approximately ten-thousand in this work). These low energy particle motions and pressure fluctuations could alternatively be captured by modifying the SPH code to output (directly) the force imparted by the fluid on each of the boundary particles as a function of time. For fluid flows dominated by energetic collisions, such as those investigated in this work, the increased number of simulated data points (i.e. time series for approximately 13\,million ice particles) would lead to significantly greater computation times than the impulse capturing method of the present work. 

We note that the seismic signals produced using this modelling framework may have some dependence on the chosen boundary implementation, for example the choice of a free slip (as in the work) or no slip boundary. While such choices may have limited influence on the bulk flow \citep[see][]{Valizadeh2015}, the behaviour of particles at the fluid-solid interface could have greater sensitivity to such choices. Further investigation of whether the choice of SPH parameters might introduce any artifacts due to the computation scheme, aside from the characteristics expected from true water, is the subject of future work.

\subsection{Displacement and velocity time series}
\label{sec:Displacement and velocity time series}

The method presented in Sections \ref{sec:Collision as a damped harmonic oscillator} and \ref{sec:Wave generation at a water-ice interface} generates an acceleration time series, while the displacement and velocity time series' may be more readily comparable to observed short period or broadband seismic records. We caution against a simple numerical integration of the acceleration time series (as one might carry out using an observational seismology library) because the time averaging is occurring over a window greater than the frequency of the signal, instead suggesting an approach as follows.

The amplitudes of the displacement and velocity time series are highly dependent on the assumed collision duration $\tau$ (Section \ref{sec:Droplet collisions with the water-ice interface}). That is, taking the time-average of the displacement for the $n$th fluid particle collision (Equation \ref{eta_x}) over a temporal sample of duration $\delta t \gg \tau_n$ yields:
\begin{equation}
\begin{split}
\bar{\boldsymbol{\eta}}_{n}(t) &= \frac{1}{\delta t}\int_t^{t+\delta t} \boldsymbol{\eta}_{n}(t^*) dt^* \\
&= \frac{1}{\delta t}\int_{t_n}^{t_n+\tau_n} \boldsymbol{\eta}_{n}(t^*) dt^* \\
&\approx \frac{\pi \tau_n}{2 \delta t} \norm{\boldsymbol{\eta}_{0,n}} \,,
\end{split}
\label{etaseries}
\end{equation}
where the second line (of Equation \ref{etaseries}), for the purposes of this argument, assumes the collision is completely contained within a single temporal sample, and the final approximate equality yields the average displacement over the collision ignoring damping. However, $\norm{\boldsymbol{\eta}_{0,n}} = \tau_n \norm{\boldsymbol{v}_{0,n} - \boldsymbol{v}_{\!\;\!\star,n}}/\pi$ and thus the time-average of the displacement is given by,
\begin{equation}
\bar{\boldsymbol{\eta}}_{n}(t) \approx \frac{\tau_n^2}{2 \delta t} \norm{\boldsymbol{v}_{0,n} - \boldsymbol{v}_{\!\;\!\star,n}} \,.
\end{equation}
The displacement time series thereby scales in proportion to the square of the collision duration. Similarly, we can show that the velocity time series scales linearly with the collision duration. However, the constant factor ensures that although the magnitude of the displacement and velocity will be unknown, other characteristics of their waveforms will be accurate.

\begin{figure*}
\centerline{\includegraphics[width=0.95\textwidth,trim={57.5 12.5 87.5 45},clip]{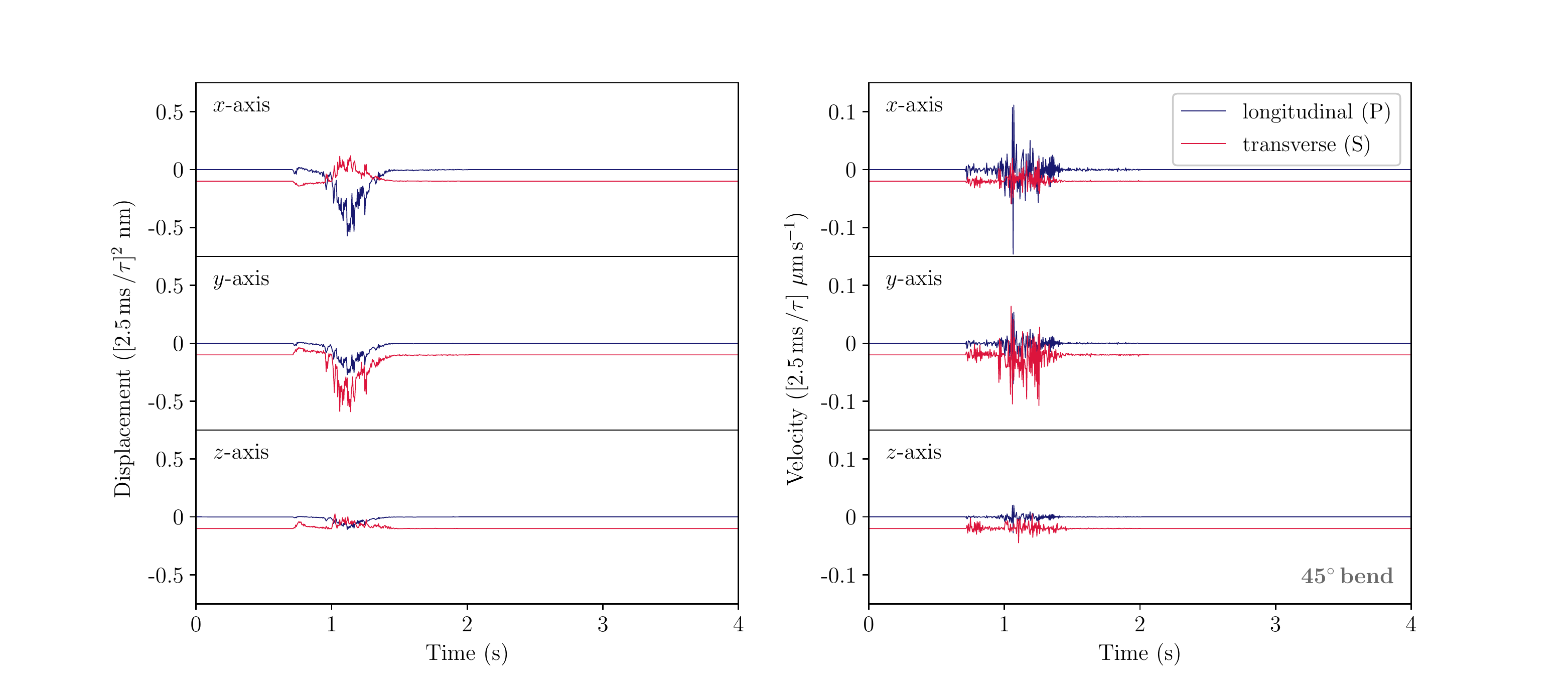}}
\caption{\textit{Left:} Displacement for synthetic seismic waves generated by water flow in the $45^\circ$ bend geometry channel measured at a seismometer located at $(x, y, z) = (10, 5, 2)$\,m. The longitudinal (P) component of the wave is shown in blue and the transverse (S) component in red, offset by $-0.1\rm\, nm$ for clarity. \textit{Right:} Velocity for the same channel geometry and seismometer location. The transverse wave is offset by $-0.02\rm\, \mu m\, s^{-1}$ for clarity. The displacement and velocity both scale with the assumed collision duration (unlike the acceleration); the units for both quantities are therefore presented as a function of the collision duration $\tau$.}
\label{fig:waveform_dv}
\end{figure*}

The character of synthetic three-component displacement and velocity waveform data generated for the 45$^\circ$ bend supraglacial channel geometry is shown (Figure \ref{fig:waveform_dv}).  These velocity time series are comparable to typical recorded broadband seismic records.

\subsection{Waveform character}

\begin{figure*}
\centerline{\includegraphics[width=0.95\textwidth,trim={57.5 12.5 87.5 45},clip]{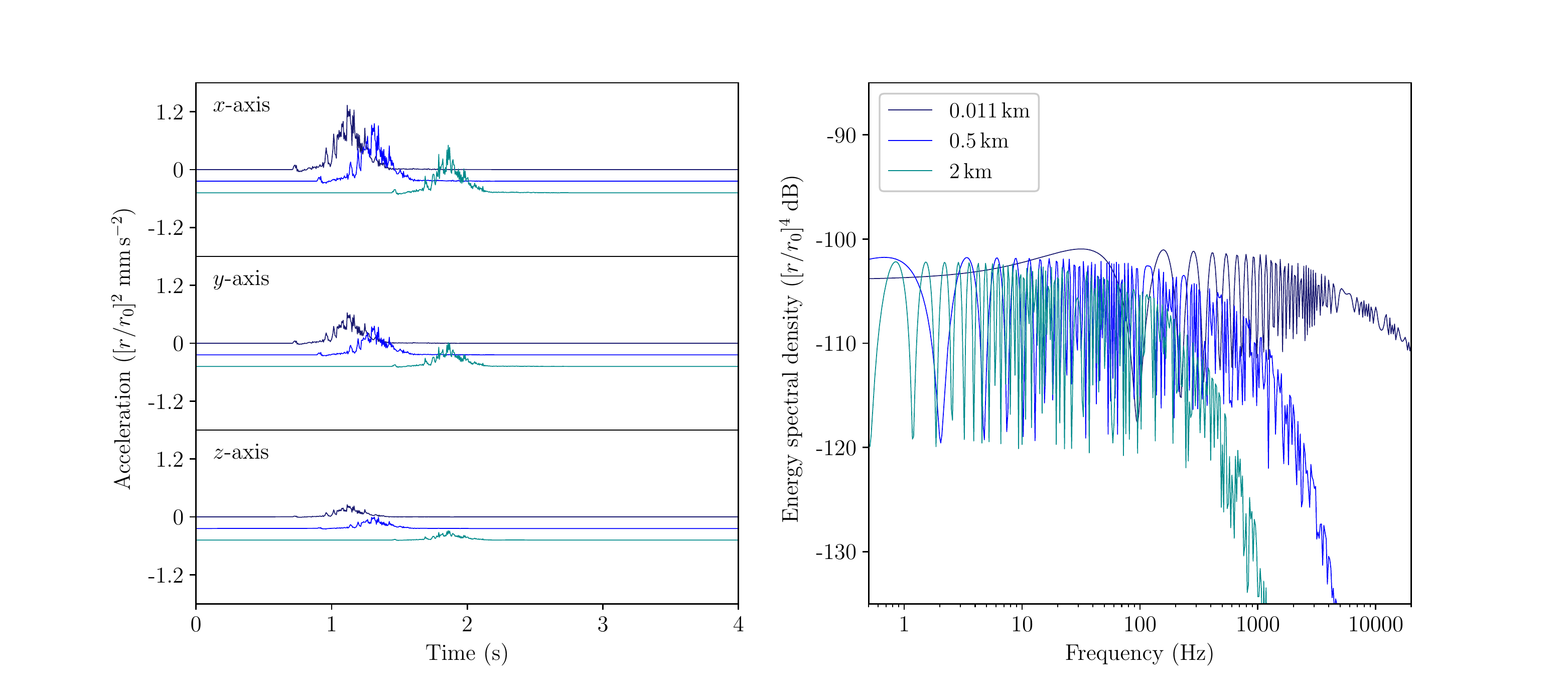}\vspace{6pt}}
\centerline{\includegraphics[width=0.95\textwidth,trim={57.5 12.5 87.5 45},clip]{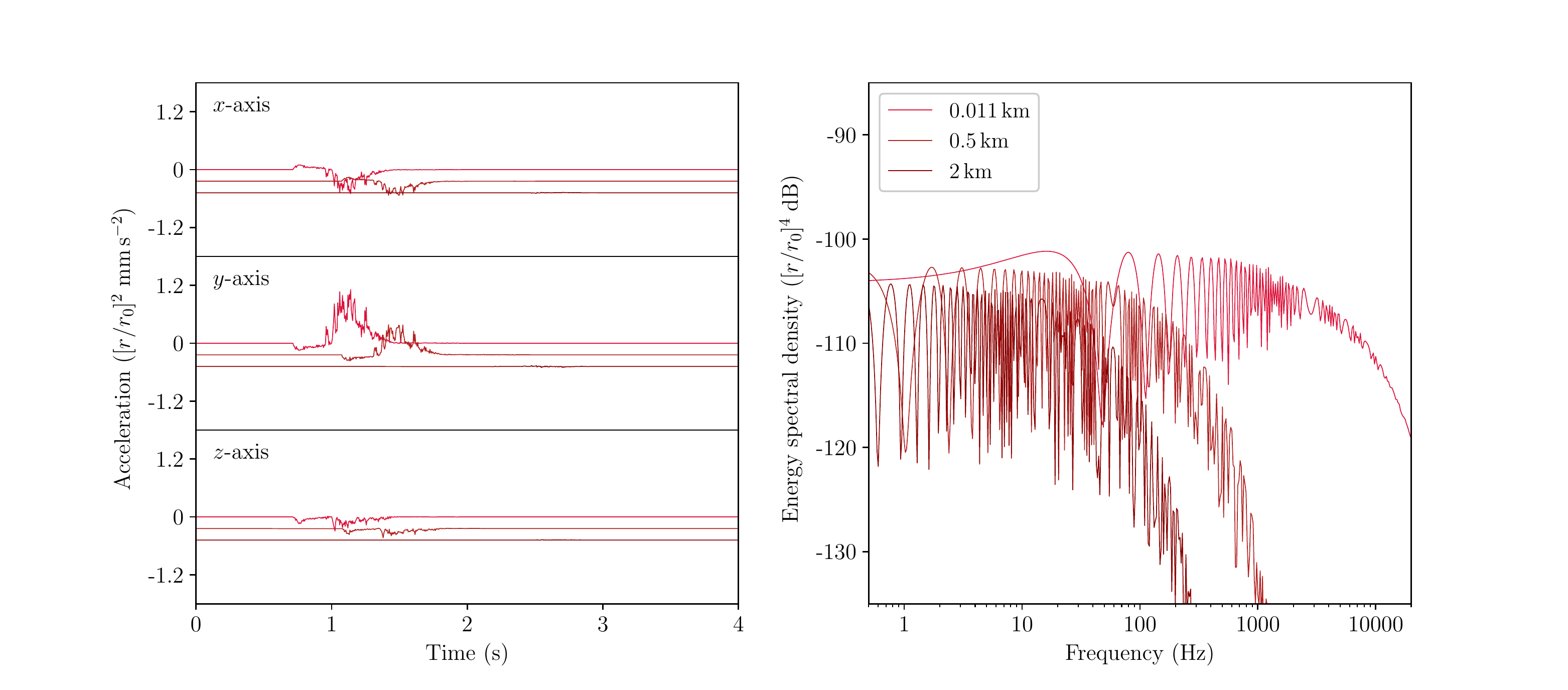}}
\caption{Waveform character (acceleration amplitude) and energy spectral density as a function of distance from the $45^\circ$ bend geometry supraglacial channel; the seismometers are located in the $(x,y,z) = (10, 5, 2)$ direction at distances of 11.35, 500 and 2000\,m from the mid-channel reference point, at an azimuthal angle of $\phi = 26.6^\circ$ and inclination angle of $\theta = 11.3^\circ$ (cf. Section \ref{sec:Synthetic time series}). The acceleration amplitude and energy spectral density are presented with the $1/r^2$ attenuation term removed (scaled to $r_0 = 11.35$\,m) for ease of comparison; attenuation and weak dispersion in the medium are included. \textit{Top:} longitudinal (P) wave and \textit{Bottom:} transverse (S) wave.}
\label{fig:atten}
\end{figure*}

\begin{figure*}
\centerline{\includegraphics[width=0.95\textwidth,trim={105 12.5 85 45},clip]{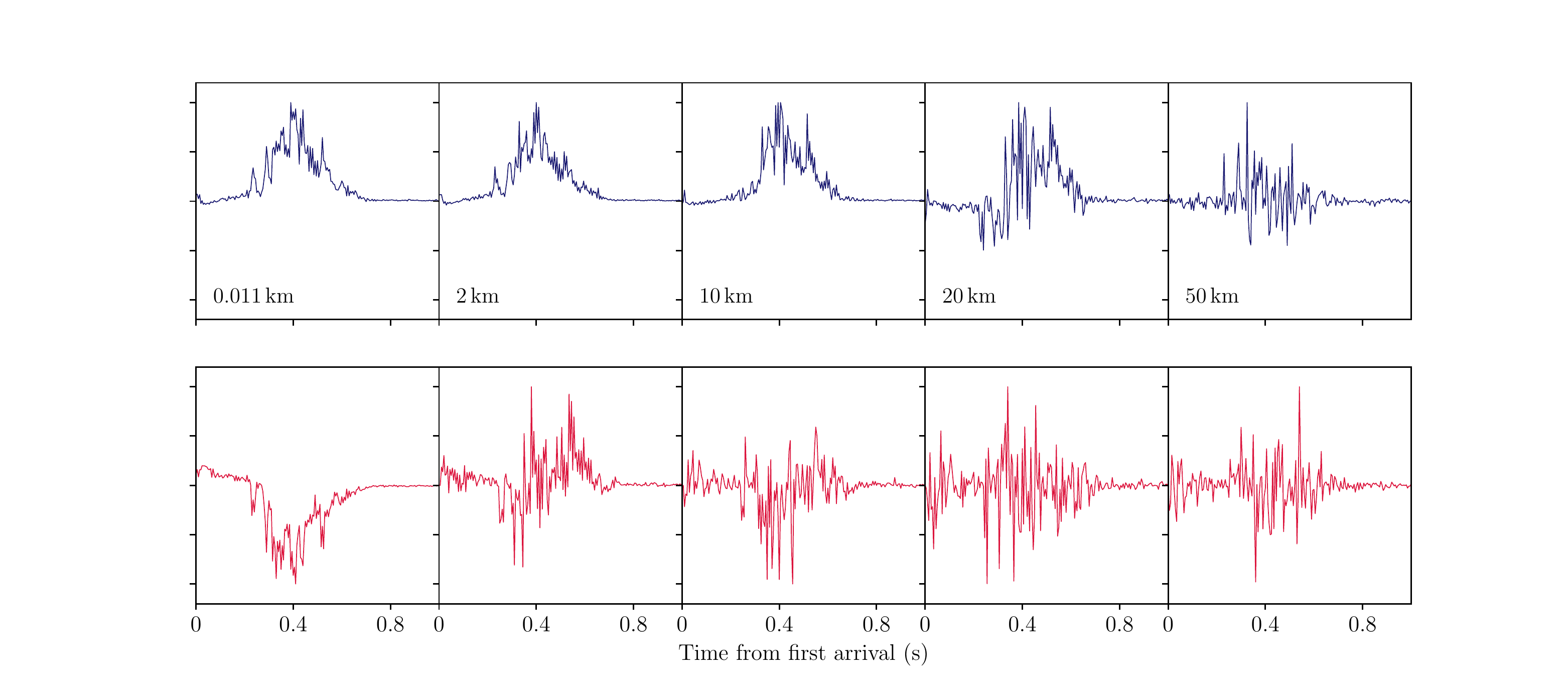}}
\caption{Weak dispersion of waveform from the $45^\circ$ bend geometry supraglacial channel; the seismometers are located in the $(x,y,z) = (10, 5, 2)$ direction at distances of 11.35\,m, 2, 10, 20 and 50\,km. The waveform acceleration amplitude ($x$-component) is scaled to unity and the time axis shifted such that the time of first arrival is zero. \textit{Top:} longitudinal (P) wave and \textit{Bottom:} transverse (S) wave.}
\label{fig:disp}
\end{figure*}

\subsubsection{Change of character with channel geometry}

Channel geometry has a significant effect on the appearance of the time series, as previously described (Section \ref{sec:Synthetic time series}). Sharp bends in the channels result in more highly peaked acceleration envelope shapes in comparison to the s-shaped curved channels (Figure \ref{fig:waveform}), while the differing angle between successive sections of the channels leads to a higher second peak in the case of both the $45^\circ$ bend and s-curve. Thus, this example illustrates the range of waveform attributes that should be considered in analysing signals from a seismometer location relatively close to a supraglacial channel, inset slightly into the surface of a glacier. Meanders feature commonly in supraglacial streams, and the development of sharp bends and small plunge pools is also plausible \citep{knighton1981channel, karlstrom2013meander, stgermain2016thedevelopment}.

\subsubsection{Change of character with distance}

The character of the seismic body wave changes with distance due to both attenuation and weak dispersion. The increasing energy loss in anelastic ice owing to the attenuation of the high-frequency spectral content (Section \ref{sec:Synthetic time series}) is captured as a function of distance for the 45$^\circ$ bend supraglacial channel (Figure \ref{fig:atten}) at three seismometer locations at increasing distances from the channel. The longitudinal (P) wave acceleration amplitude decreases by a factor of three over this distance range (Figure \ref{fig:atten}, top-left), whereas the transverse (S) wave decreases in amplitude by an order of magnitude (Figure \ref{fig:atten}, bottom-left). The energy spectral density for these seismic waves is shown in Figure \ref{fig:atten} (right column); the high-frequency cut-off in the energy spectral density function shifts to progressively lower frequencies with distance for the supraglacial channel for both wave types. However, the S wave energy is additionally reduced by a factor of approximately two, even at the lowest frequencies, explaining the greater drop in acceleration amplitude.  The character of the observed seismic waveform changes further with greater distance due to dispersion (Figure \ref{fig:disp}), although in practice the waveform would probably be modified by structures in the ice or other medium as the seismic signal travels over larger distances. Regardless, the results of our model show the waveform character may change significantly over a distance of just 2\,km for seismometers at angular locations around the source dominated by the S wave signal.

\subsubsection{Ongoing Work}

This contribution presents a two-part modelling framework, combining SPH simulations and seismic wave propagation, applied to a simple set of channel geometries representing narrow, supraglacial flows. The approach has the potential to be applied in ongoing work to investigate signals arising from further supraglacial flow geometries, and also englacial and subglacial channel flows. As the geometries and associated flows become more complex, the flexible nature of the framework will be beneficial, allowing the fluid motion that initiates the seismic disturbance to be tailored to the situation. As noted previously, resonance and sediment transport processes are also of relevance in some situations and could be included in our modelling as refinements to the SPH simulations or by utilising findings from previous studies of these mechanisms.  Where there is a case for understanding more subtle signal generation processes such as when pressure fluctuations in the fluid matter, it would be possible to use the SPH part of the framework to track fine details of the water-ice interaction at cost of increased computation time.

A wider goal of the simulations enabled by the framework is to inform the exploration of glacier processes through waveform attributes. The framework should therefore enable insights to be drawn through comparison with observed patterns of waveform attributes extracted through approaches such as unsupervised learning applied to seismic records from instruments deployed in glacier environments.

\section{Conclusions}
\label{sec:conclusions}

In this contribution, we have introduced a modelling framework tailored to progressing the understanding of body wave signals produced by moving water in glacier environments. The framework is in two parts: (1) simulation of water movement, using a smoothed particle hydrodynamic (SPH) approach, constrained in channels or similar features; (2) modelling of particle motion due to the impact of the moving water on the side of the channel, and subsequent wave propagation.

% Para 1 - Modelling framework 
The SPH part of the modelling framework can be configured to simulate water motion in a wide variety of channel or other geometries, and effectively creates a reusable dataset of collisions. These collisions are then utilised by the second part of the framework. This allows for computational efficiency in, for example, exploring the nature of waveforms at different locations across the ice. The framework has considerable flexibility as, in particular, the wave propagation can make use of standard outputs from alternative packages to that used in this work to generate the input hydrodynamic simulation.

% Para 2 - Strengths of the approach
The seismic wave modelling is formulated to yield acceleration waveforms that are invariant to the temporal and spatial resolution of the hydrodynamic simulation, and the assumed time scale of the collision. We further make use of the Fourier spectrum of the time-varying behaviour of identified collisions to enable weak dispersion and attenuation to be considered. The framework as presented is well-suited to energetic collisions (e.g. water moving under gravity) but is less-suited to lower energy interactions (e.g. pressure waves or shallow angle collisions) as we filter out these fluid particle oscillations for computational efficiency. We have proposed minor modifications to our algorithm to capture signals due to such flows if required in future work, albeit with potentially greatly increased run times.

% Para 3 - Findings
We find that the character of the simulated body wave envelope or the waveform itself (for example, number of peaks, spiked or undulating, high or moderate peak amplitude, P amplitude:S amplitude ratio) is dependent on details of the channel geometry. The angle of collision with interface has a particularly significant effect. Changes in the waveform due to propagation effects are also captured such that the resulting signal simulation for seismometer locations at varying distance can now inform comparisons, in ongoing work, with observed signals.
% Para 4 - Applications
The aim of the framework is to progress the understanding of the character of seismic signals generated in the glacier environment and reasons for the variability of waveform attributes. We anticipate that the computational capability will inform the identification of processes and process change in the future monitoring of remote glaciers.
%\clearpage

\section*{Acknowledgments}
This research was supported under Australian Research Council's Discovery Project DP210100834, with additional contributions from the Australian Government's Antarctic Science Collaboration Initiative program (SC), the ARC SRI Australian Centre for Excellence in Antarctic Sciences and a University of Tasmania Scholarship (JCM).

\section*{\phantomsection
DATA AVAILABILITY}
\label{sec:Data Availability}

The authors confirm that the data supporting the findings of this study are available within the article, and the relevant code and generated data products are publicly available on GitHub (\href{https://github.com/rossjturner/seismicsph}{github.com/rossjturner/seismicsph}).
\vfill

\bibliographystyle{apalike}
\bibliography{seismic_hydro_sims.bib}

\appendix
\section{Seismic waves in glacier ice}
\label{sec:Seismic waves in glacier ice}

The anelasticity of glacial ice is explained a combination of: (1) the interatomic potentials in the hexagonal crystal structure on atomic scales; or (2) the granular nature of ice on macroscopic scales. The granular structure results from the nucleation sites associated with ice formation, leading to weak points in the crystal lattice at the grain boundaries. The slipping of ice crystals along these boundaries dominates the elastic properties of glacial ice at seismic wave frequencies, and thus is the focus of this section. The literature largely considers the response of ice at higher frequencies as we discuss in detail in Section \ref{sec:Seismic waves}.

\subsection{Young's modulus of glacier ice}

\citet{Traetteberg+1975} constrain the stress-strain relationship in both artificial and naturally formed ice at $-10^\circ\rm C$ under the application of a compressive load. The Young's modulus (derived from stress/strain) is strongly correlated with the relaxation time of the applied stress in the material; their measured relaxation times correspond to the frequency range 0.0001-1.5\,Hz. Meanwhile, \citet{Nakaya+1958} measure the Young's modulus of tunnel ice from Tuto, Greenland at a range of temperatures by measuring the resonant frequency of the polycrystaline ice at 200-600\,Hz. The elastic properties at MHz-GHz frequencies are dominated by the interatomic potentials within ice crystals; e.g. \citet{Gammon+1983} use Brillouin spectroscopy (at 10\,GHz) to measure the Young's modulus of glacier ice at $-16^\circ\rm C$ as $E = 9.332\rm\, GPa$ (equivalent to $E = 9.254\rm\, GPa$ at $-10^\circ\rm C$). 

\begin{figure}
\centerline{\includegraphics[width=\columnwidth,trim={10 15 30 45},clip]{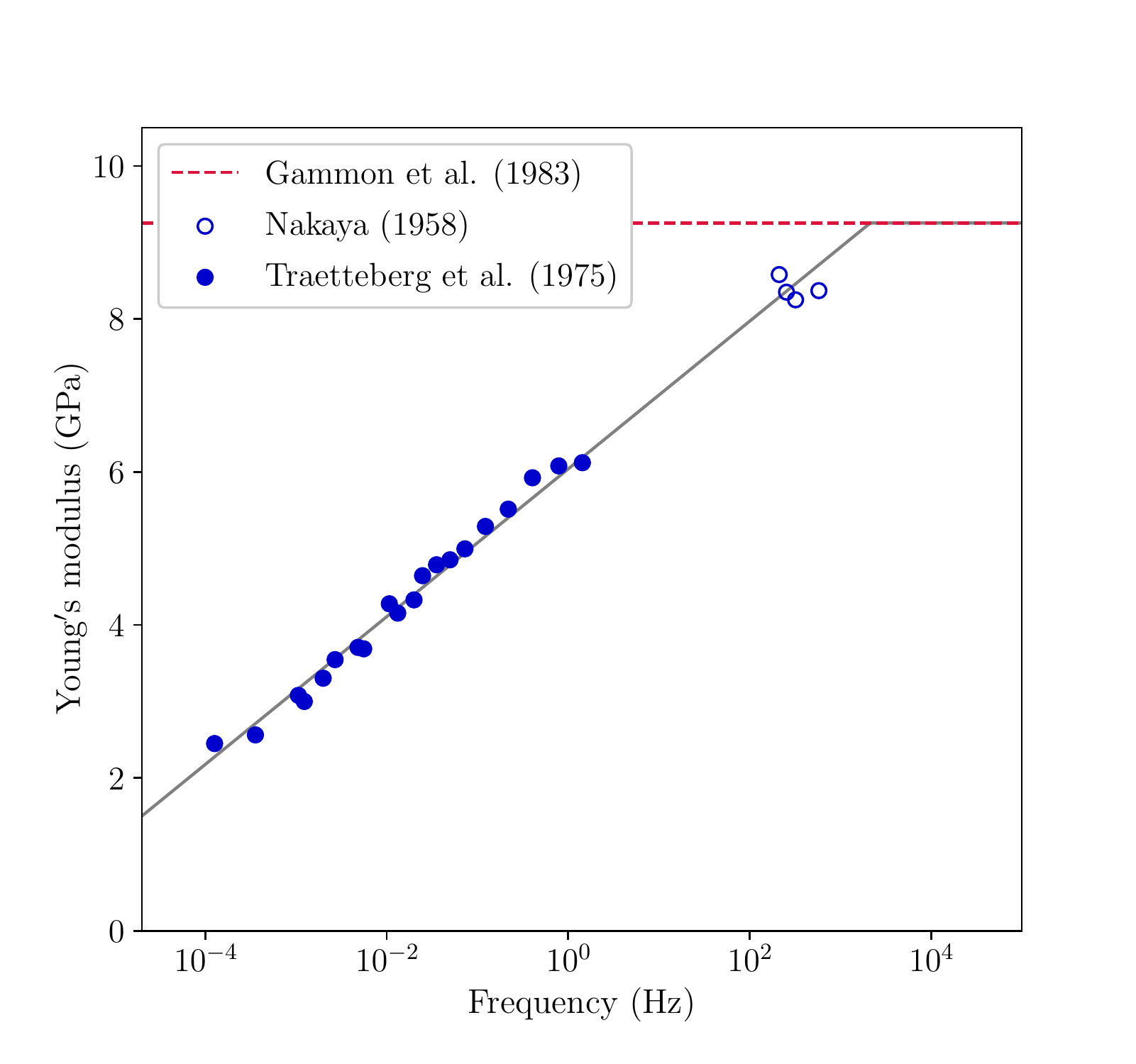}}
\caption{Frequency dependence of the Young's modulus in glacial ice at a reference temperature of $-10^\circ \rm C$. Results from \citet{Nakaya+1958} and \citet{Traetteberg+1975} are shown by the filled/unfilled circles. The Brillouin spectroscopy measurement \citet{Gammon+1983}, shown by the dashed red line, provides an approximate upper bound on the Young's modulus. The grey line is the fit of a conditional linear function to these measurements over the shown frequency range.}
\label{fig:elastic_moduls}
\end{figure}

We model the frequency dependence of the Young's modulus at a reference temperature of $T_\text{ref} = -10^\circ \rm C$ using a conditional linear function. That is,
\begin{equation}
E(\xi,T_\text{ref}) = \begin{cases}
6.041 + 0.966{\,\log_{10}}{\xi} \rm\; GPa,  \ & \xi < 2120\rm\, Hz \\
9.254 \rm\; GPa,  \ & \xi \geqslant 2120\rm\, Hz
\end{cases}\,,
\end{equation}
where the frequency $\xi$ is in Hertz; this expression yields non-positive elastic moduli for $\xi < 6\times 10^{-7}\rm\, Hz$, and thus should not be extrapolated significantly beyond seismic wave frequencies. The fit and literature measurements are shown in Figure \ref{fig:elastic_moduls}.

The temperature dependence of the Young's modulus is modelled following the method of \citet[][and subsequently \citealt{Gammon+1983}]{Dantl+1969} for measurements of the elastic properties of ice at high frequencies. However, we instead fit the functional form of their proposed temperature correction to the measurements of \citet{Traetteberg+1975} for naturally formed ice over the $-39.5^\circ \rm C$ to $-10^\circ \rm C$ temperature range. That is,
\begin{equation}
E(\xi,T) = E(\xi,T_\text{ref})\frac{1 - aT}{1 - aT_\text{ref}} \,,
\end{equation}
where the temperature $T$ is in degrees Celsius and $a = 0.0516\rm\,/ ^\circ C$ (cf. $a = 1.418\times10^{-3}\rm\,/ ^\circ C$ for the fit at high frequencies).

% Para 5, More material properties of ice
\subsection{Dispersion of seismic waves in glacier ice}
\label{sec:Frequency dependent wave propagation in ice}

\citet{Kohnen+1973} used seismic refraction measurements near Byrd Station, West Antarctic to constrain the phase velocity of transverse (S) and longitudinal (P) seismic waves propagating in glacial ice as a function of density. The ratio of these velocities is used the derive Poisson's ratio at seismic wave frequencies (cf. Equations \ref{a1} and \ref{a2}) as $\sigma = 0.33$. The value of Poisson's ratio is comparable for all ice densities considered in the analysis of \citet{Kohnen+1973}, except those in the top few metres of the firn layer. We therefore assume a constant value for this parameter, noting that our modelling is restricted to at least somewhat comparable temperatures and wave frequencies. 

The dispersion of seismic waves is parameterised in this work in terms of the wavenumber of the S and P waves. These are related to Poisson's ratio and Young's modulus as follows:
\begin{subequations}
\begin{equation}
k_{p}(\xi, T) = 2\pi \xi \bigg[\frac{1.482E(\xi, T)}{\varrho(T)} \bigg]^{-1/2}
\label{wavenumber1}
\end{equation}
\vspace{-8pt}
\begin{equation}
k_{s}(\xi, T) = 2\pi \xi \bigg[\frac{0.376E(\xi, T)}{\varrho(T)} \bigg]^{-1/2} \,,
\label{wavenumber2}
\end{equation}
\end{subequations}
where $\varrho(T = 0^\circ{\rm C}) = 916\rm\, kg\, m^{-3}$ is the density of ice at atmospheric pressure.
The derivative of the wavenumber is required for the Taylor series approximation of the dispersion relationship in Equation \ref{k_x}. That is,
\begin{subequations}
\begin{equation}
\frac{\partial k_{p}(\xi, T)}{\partial \xi} = \bigg(2\pi - \frac{\pi \xi}{E(\xi)} \frac{\partial E(\xi)}{\partial \xi} \bigg) \bigg[\frac{1.482E(\xi, T)}{\varrho(T)} \bigg]^{-1/2} 
\label{wavenumberd1}
\end{equation}
\vspace{-8pt}
\begin{equation}
\frac{\partial k_{s}(\xi, T)}{\partial \xi} = \bigg(2\pi - \frac{\pi \xi}{E(\xi)} \frac{\partial E(\xi)}{\partial \xi} \bigg) \bigg[\frac{0.376E(\xi, T)}{\varrho(T)} \bigg]^{-1/2}  ,
\label{wavenumberd2}
\end{equation}
\end{subequations}
where the derivative of the frequency-dependent elastic modulus at the reference temperature, $T_\text{ref} = -10^\circ\rm C$, is given by:
\begin{equation}
 \frac{\partial E(\xi, T_\text{ref})}{\partial\xi} = \begin{cases}
\frac{0.419}{\xi} \rm\; GPa\,s,  \ & \xi < 2120\rm\, Hz \\
0 \rm\; GPa\,s ,  \ & \xi \geqslant 2120\rm\, Hz
\end{cases}\,.
\end{equation}

% Para 6, Attenuation
\subsection{Attenuation of seismic waves in glacier ice}
\label{sec:attenuation}

\begin{figure}
\centerline{\includegraphics[width=\columnwidth,trim={10 15 30 45},clip]{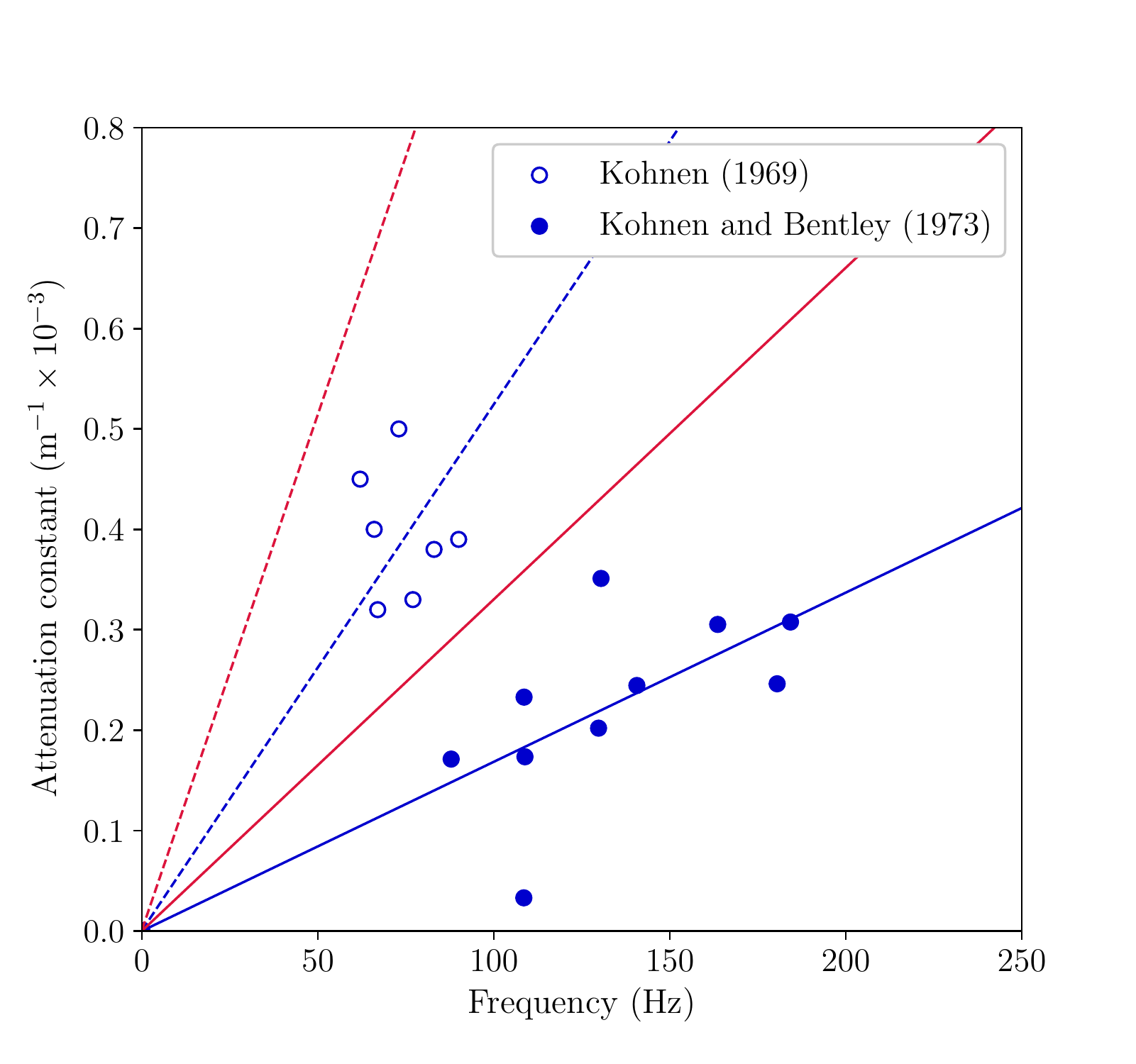}}
\caption{Frequency dependence of the longitudinal wave attenuation constant in ice. Results from \citet{Kohnen+1971} are shown by filled circles and earlier measurements by \citet{Kohnen+1969} with unfilled circles; these are fitted by linear functions with blue solid and dashed lines respectively. The relationships for the transverse wave attenuation constant corresponding to the measurements by these authors are shown in red.}
\label{fig:attenuation}
\end{figure}

The attenuation of seismic waves in granular media is found to be independent of the wave amplitude, linearly proportional to the wave frequency, and dependent on the coefficients of static and kinetic friction at the boundary of ice crystals \citep{Attewell+1966, White+1966}. The attenuation of P waves was investigated by \citet{Kohnen+1971} using seismic refraction measurements near Byrd Station (as part of a series of publications, including that discussed in the previous section). Their measurements (in the 80-180\,Hz waveband) show a linear relationship between the attenuation constant and wave frequency of the form $\alpha_p(\xi) = (0.15\xi + 3)\times 10^{-5}\rm\, m^{-1}$. However, previous monochromatic measurements on the Antarctic and Greenland ice sheets \citep[e.g.][]{Robin+1958, Kohnen+1969} found an attenuation constant a factor of approximately two greater than predicted by this relationship. We therefore also consider a linear relationship scaled to fit measurements by \citet{Kohnen+1969} across five stations. These two fits are shown in Figure \ref{fig:attenuation} for comparison. We do not consider the temperature dependence on the attenuation constant due to the large variance between measurements at just a single temperature.

Meanwhile, the lower phase velocity of the S wave (relative to the P wave), and thus greater number of oscillations per unit length, results in an increased interaction with each ice grain boundary ($\propto 1/v$) and thus more rapid attenuation of the wave amplitude. 
The ratio of the phase velocities for the S and P waves are independent of frequency (cf. Equations \ref{wavenumber1} and \ref{wavenumber2}); we therefore expect the attenuation constant for S waves in glacial ice to be greater than the measured P wave value by $\alpha_s/\alpha_p = 1.99$. Regardless, given the attenuation constant is of order $10^{-4}\rm\, m^{-1}$ for typical seismic wave frequencies, we do not expect this approximation to meaningfully affect our results for distances up to at least 10\,km. The frequency-dependent relationships for the attenuation constant of the S waves are shown in Figure \ref{fig:attenuation} in comparison to the P waves measurements. 

The expressions used in this work for the frequency-dependent  attenuation constant of the S and P waves are summarised below. That is,
\begin{subequations}
\begin{equation}
\alpha_p(\xi) = (0.15\xi + 3)\times 10^{-5}\rm\, m^{-1}
\label{alpha1}
\end{equation}
\vspace{-16pt}
\begin{equation}
\alpha_s(\xi) = (0.30\xi + 5.96)\times 10^{-5}\rm\, m^{-1} \,,
\label{alpha2}
\end{equation}
\end{subequations}
whilst their derivative required in the Taylor series expansion for the inverse-Fourier transform are given by, 
\begin{subequations}
\begin{equation}
\frac{\partial \alpha_p(\xi)}{\partial \xi} = 0.15\times 10^{-5}\rm\, m^{-1}\, s
\label{alphad1}
\end{equation}
\vspace{-8pt}
\begin{equation}
\frac{\partial \alpha_s(\xi)}{\partial \xi} = 0.309\times 10^{-5}\rm\, m^{-1}\, s \,.
\label{alphad2}
\end{equation}
\end{subequations}
These relationships are mathematically valid for any frequency range, though of course should be applied close to the 80-180\,Hz range of measurements used to derive them.

\bsp % ``This paper has been produced using the Blackwell
     %   Publishing GJI \LaTeXe\ class file.''

\label{lastpage}

\end{document}